\documentclass{aa}  

\usepackage{graphicx}
\usepackage{txfonts}
\usepackage{amsmath}
\usepackage{soul}

\newcommand{\macc}{$\dot{M}_{\rm acc}$}

\usepackage{color}

\begin{document} 

\title{Accretion variability from minutes to decade timescales\\ in the classical T Tauri star CR Cha\thanks{Based on observations collected at the European Southern Observatory under ESO programs 2103.C-5025, 0103.A-9008, and 0100.C-0708(A)}}

\author{G. Zsidi \inst{1,2,4}
\and
C. F. Manara\inst{1} 
\and
\'A. K\'osp\'al \inst{2,3,4}
\and
G. A. J. Hussain\inst{1,5}
\and
P. \'Abrah\'am \inst{2,4}
\and \\
E. Alecian \inst{6}
\and
A. B\'odi \inst{2,7}
\and
A. P\'al \inst{2}
\and
P.~Sarkis \inst{3}
}

\institute{European Southern Observatory, Karl-Schwarzschild-Strasse 2, 85748 Garching bei M\"unchen, Germany \\
e-mail: zsidi.gabriella@csfk.org
\and
Konkoly Observatory, Research Centre for Astronomy and Earth Sciences, E\"otv\"os Lor\'and Research Network (ELKH), Konkoly-Thege Mikl\'os \'ut 15-17, 1121 Budapest, Hungary
\and
Max Planck Institute for Astronomy, K\"onigstuhl 17, 69117 Heidelberg, Germany
\and
ELTE E\"otv\"os Lor\'and University, Institute of Physics, P\'azm\'any P\'eter s\'et\'any 1/A, 1117 Budapest, Hungary
\and 
European Space Agency (ESA), European Space Research and Technology Centre (ESTEC), Keplerlaan 1, 2201 AZ Noordwijk, The Netherlands
\and
Univ. Grenoble Alpes, CNRS, IPAG, F-38000 Grenoble, France
\and
MTA CSFK Lend\"ulet Near-Field Cosmology Research Group, 1121, Budapest, Konkoly Thege Mikl\'os \'ut 15-17, Hungary
}


 
  \abstract
   {Classical T Tauri stars are pre-main sequence stars that are surrounded by a circumstellar disk from which they are accreting material. The mass accretion process is essential in the formation of Sun-like stars. Although often described with simple and static models, the accretion process is inherently time variable. }
   {The aim of our study is to examine the accretion process of the low-mass young stellar object CR Cha on a wide range of timescales from minutes to a decade by analyzing both photometric and spectroscopic observations from 2006, 2018, and 2019.}
   {We carried out period analysis on the light curves of CR Cha from the TESS mission and the ASAS-SN and the ASAS-3 databases. We studied the color variations of the system using $I,J,H,K$-band photometry obtained contemporaneously with the TESS observing window. We analyzed the amplitude, timescale, and the morphology of the accretion tracers found in a series of high-resolution spectra obtained in 2006 with the AAT/UCLES, in 2018 with the HARPS, and in 2019 with the ESPRESSO and the FEROS spectrographs.}
   {All photometric data reveal periodic variations compatible with a 2.327 days rotational period. In addition, the ASAS-SN and ASAS-3 data hint at a long-term brightening by 0.2 mag, between 2001 and 2008, and of slightly less than 0.1 mag in the 2015 -- 2018 period. The near-infrared photometry indicate a short-term brightening trend during the observations in 2019. The corresponding color variations can be explained by either
   changing accretion rate or changes in the inner disk structure. The H$\alpha$ line profile variability suggests that the amplitude variations of the central peak, likely due to accretion, are the most significant on daily/hourly timescales. On yearly timescales, the line morphology also changes significantly. }
   {The photometric variability shows that the $\sim$2.3-day period is stable in the system over decades. Our results show that the amplitude of the variations in the H$\alpha$ emission increases on timescales from hours to days/weeks, after which it stays similar even when looking at decadal timescales. On the other hand, we found significant morphological variations on yearly/decadal timescales, indicating that the different physical mechanisms responsible for the line profile changes, such as accretion or wind, are present to varying degrees at different times.}


   \maketitle
%

\section{Introduction}
Classical T Tauri stars are low-mass stars at an early stage of their evolution. They are surrounded by a circumstellar disk made of gas and dust, from which they are accreting material. The current paradigm of the accretion process is best described by the magnetospheric accretion model \citep{hartmann2016}. According to this model, the innermost part of the disk is truncated by the stellar magnetic field at a few stellar radii. From this truncation radius, the disk material is channeled along the magnetic field lines from the inner disk onto the stellar surface \citep{bouvier07}. This process is best probed with spectroscopy.

Photometric and spectroscopic observations of classical T Tauri stars show that variability is a general characteristic of these targets. 
These young stars show both rapid and slow variations which appear either irregular or, in many cases, periodic or quasi-periodic \citep{herbst1994}. Different physical processes can give origin to these variations. These include variable accretion, rotational modulation, obscuration by a dust cloud between the observer and the source, or rapid structural changes in the inner disk \citep{cody2014}. Studying such variations in photometric and spectroscopic data of young stars is key to shedding light on the variability of the accretion process, and on the evolution over time of the stellar photosphere and their inner disks. 

The spectra of T Tauri stars display some peculiar features \citep{herbig1962}. Photospheric lines are generally shallow due to the veiling of the spectra caused by excess continuum emission arising from the accretion process \citep[e.g.,][]{calvet98,manara13}. As accretion is variable also on short timescales of a day or less, the veiling of photospheric lines can also vary on similar timescales. 
The spectra of T Tauri stars are also characterized by the presence of several strong emission lines, which are also highly variable as they are thought to originate in accretion columns \citep[e.g.,][]{muzerolle98}. Typical accretion tracers in the optical wavelength range are the hydrogen Balmer lines, He~I, Ca~II or Na~I \citep[e.g.,][]{alcala14,hartmann2016}. The flux and shape of these lines depend on the geometry of the accretion process, the physical properties in the magnetospheric region, and on the mass accretion rate (\macc) on the star. 
The large line widths of some emission lines, whose maximum velocities are roughly consistent with the free-fall velocities, indicate that they must form in the magnetosphere. Certain emission lines -- such as the Na~I doublet, the H$\alpha$, and the He~I lines -- exhibit in many cases redshifted absorption. These are interpreted as signature of the magnetospheric accretion infall \citep[e.g.,][]{hartmann2016}.  
However, this signature could be less evident in some lines, especially in the hydrogen Balmer series. 
Indeed, while in general asymmetries and redshifted absorption components are expected in the hydrogen lines profiles, the redshifted absorption component might be completely filled in, and the overall blueward asymmetry significantly reduced as damping wing broadening can produce significant high-velocity emission in the H$\alpha$ line, and to a lesser extent in the other Balmer lines \citep{muzerolle2001}.

This study focuses on CR Cha, a classical T Tauri star with K spectral type \citep{hussain2009,manara2016}. This star is located in the Chamaeleon~I star-forming region at a distance of 184.7$\pm$0.4 pc \citep{bailer-jones2021}. It has an effective temperature of 4900 K, a stellar luminosity of $L_{\ast} = 3.3 - 3.8\ L_{\odot}$, and a rotational period of 2.3 days \citep{bouvier1986}, it shows an average brightness level of $V=11.0$ mag, with photometric variability on the scale of a few tenth of a magnitude. Considering its luminosity and temperature, the stellar mass and age reported in the literature are in the range of $M_{\ast} = 1.2 - 2 M_{\odot}$, and an age of 1-3 Myr \citep{dantona1994,siess2000, hussain2009,manara2016}. 
The star displays moderate accretion of $\dot{M}_{acc} \sim 2.8\cdot10^{-9} M_\odot$/yr \citep{manara2016,manara2019} and hosts a protoplanetary disk with $i=31^{\circ}$ inclination \citep{kim2020}. Recent ALMA observations revealed  a dust gap at $r\sim 90$ au and a ring at $r\sim 120$ au in the disk \citep{kim2020}.

CR Cha possesses a complex, multipolar magnetic field \citep{hussain2009}, which might be explained by the star having developed a radiative core. In contrast, fully convective stars have simpler, large-scale almost fully poloidal fields \citep{donati2008, reiners2009}. The complex magnetic field with weaker dipole component may allow the inner disk to penetrate closer to the star than the corotation radius. 

Here, we present a multi-epoch photometric and spectroscopic analysis of CR~Cha using data covering more than a decade with the goal of understanding the variability of the accretion process on a range of timescales. 
CR~Cha is chosen as the target of this study as it is an excellent candidate to study variability. Indeed, CR~Cha is a typical classical T~Tauri star and its magnetic field has been already mapped \citep{hussain2009}. In addition, it is known to be a single star and to have a low mass accretion rate. Both aspects make it less complicated to interpret the observed variations in the photometry and in the spectra.
The paper is structured as follows. Sect.~\ref{sect::obs} presents the three observing campaigns used in this work, and briefly describes the data reduction processes. Sect.~\ref{sect::analysis} explains how the photometric and spectroscopic data were analyzed, detailing the results of the study of both regular (periodic) and irregular photometric and spectroscopic variations. We then discuss the results in Sect.~\ref{sect::discussion} and summarize our conclusions in Sect.~\ref{sect::concluions}.


\section{Observations and data reduction}\label{sect::obs}

We monitored CR Cha in three different observing seasons. We carried out spectroscopic observations in 2006, 2018, and 2019, and we complemented our data with ground-based multifilter photometry and high-cadence space photometry. In the following sections, we describe the three observing seasons.

\subsection{2006 observing season}
Spectropolarimetric observations were made with the 3.9~m Anglo-Australian Telescope (AAT) over five consecutive nights from 2006 April 9 to 13. One to four observations were obtained per night due to varying weather conditions, which resulted in an incomplete phase coverage. A visitor polarimeter (SemelPol) was mounted at the Cassegrain focus of the telescope and was coupled with the UCLES spectrograph covering the wavelength range from 437.6 to 682 nm with spectral resolution around 70\,000. As described by \citet{hussain2009}, the spectra were reduced using the ESPRIT data reduction package, which includes bias subtraction, flat fielding, wavelength calibration, and optimal extraction of (un)polarized \'echelle spectra \citep[see][]{donati1997, donati2011, donati2014}.

$V$-band photometric observations are publicly available from the ASAS-3 catalogue \citep{pojmanski2003} for the AAT observing period. CR~Cha was observed once every 1-3 nights using a 9$^{\circ}\times$9$^{\circ}$ wide-field camera with a pixel size of 15$"$. The $V$-band magnitudes were calculated using aperture photometry through five different apertures (2 to 6 pixels) and the results are publicly available from the online catalogue\footnote{\url{http://www.astrouw.edu.pl}}. For the 2006 data, we used the dataset with the smallest aperture of 2 pixels. We note that the ASAS-3 $V$-band observations are available for the period 2001--2009. 

\subsection{2018 observing season}
Spectropolarimetric measurements were taken with the HARPS instrument on the ESO 3.6\,m telescope in La Silla (Pr.Id.0100.C-0708, PI F. Villebrun). The data were acquired using HARPS in polarimetric mode \citep{piskunov2011}, yielding spectra with a resolving power of about 105\,000 and covering wavelengths spanning from 380 to 690 nm. All spectra were recorded as sequences of four individual sub-exposures taken in different configurations of the polarimeter to allow a full circular polarisation analysis. The data reduction process was carried out using the  LIBRE-ESPRIT package \citep[][]{donati1997, hebrard2016}. In total, 24 circularly polarised sequences were acquired over six consecutive nights, from 2018 March 3 to 9, with peak S/N levels ranging from 46 to 140. For a few observations, the blue arm was not extracted due to lower S/N, therefore the whole spectrum is accessible only for 19 observations.

$V$- and $g$-band photometric observations are available from the ASAS-SN catalogue \citep{kochanek2017, shappee2014} for the HARPS observing period. 
The camera has a field of view of 4.5$^{\circ}\times$4.5$^{\circ}$ with a pixel size of 8$"$. Aperture photometry with 2 pixel aperture radius was performed to obtain the $V$- and $g$-band magnitudes and the dataset is publicly available from the online catalogue\footnote{\url{https://asas-sn.osu.edu/}}. 
Multiple measurements are available per night, however, the cadence of the ASAS-SN observations are irregular: at times, these observations were made within 1-2 hours, and at other times, they are separated by a few hours.
In total, the $V$-band observations are available for the 2014--2018 period, and $g$-band observations are carried out since the end of 2017. This means that observations made in both bands are available for the 2018 HARPS observing period.

   \begin{figure*}[ht!]
   \centering
   \includegraphics[width=\textwidth]{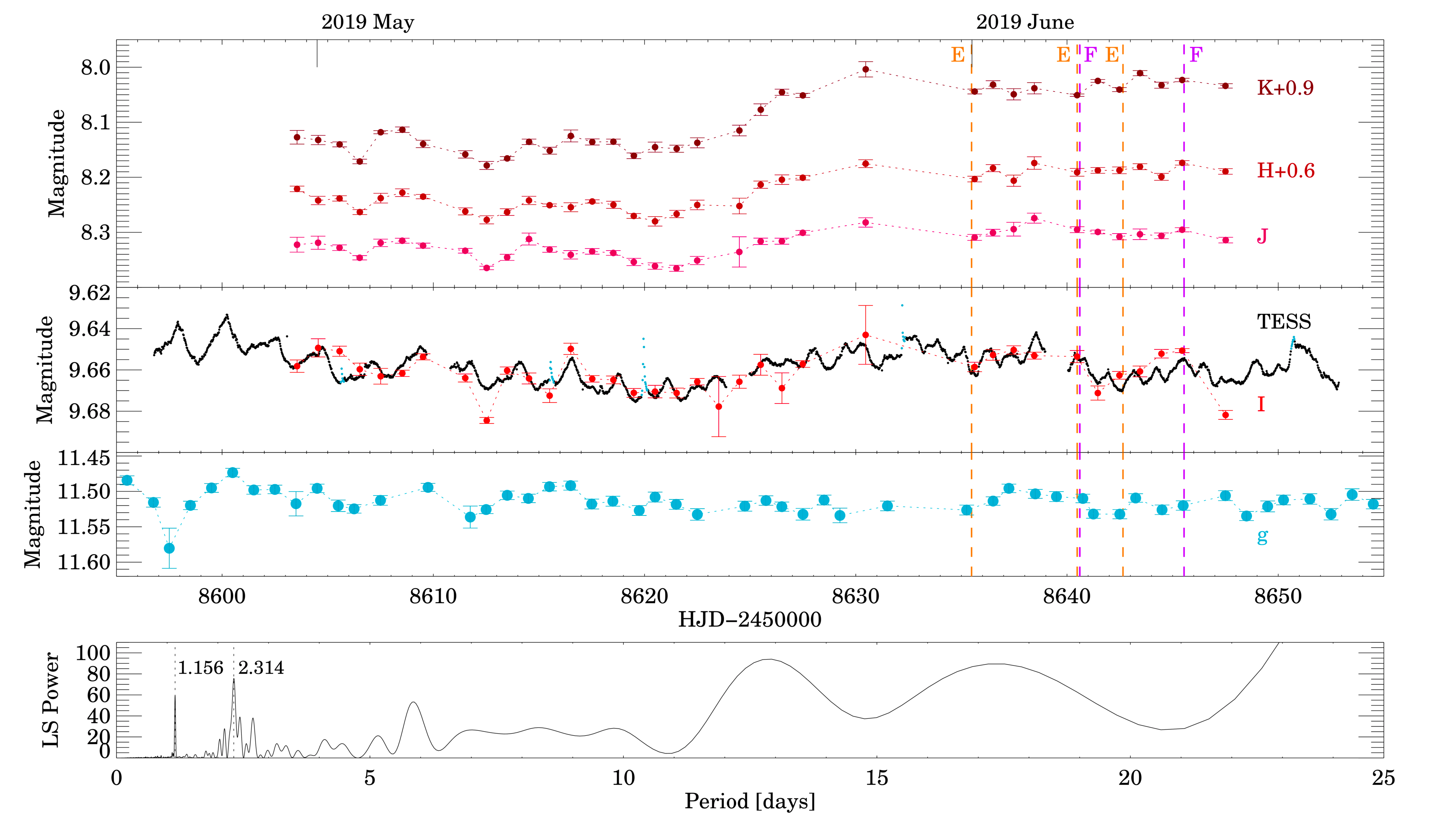}
   \caption{The TESS light curve is indicated with a black curve, and the flare-like events are coloured in blue. The ground-based $IJHK$-band measurements are shown with coloured dots, and were shifted along the $y$ axis by the values indicated in the corresponding labels in the figure. The nightly averaged ASAS-SN $g$-band observations, obtained during the TESS observing period, are plotted with blue points. The orange and purple vertical dashed lines show the epochs when the ESPRESSO (E) and FEROS (F) spectra were taken, respectively. The long tickmarks on the top of the figure indicate the beginning of each month. The bottom panel presents the Lomb-Scargle periodogram obtained using the TESS data.}
              \label{fig:tess_lc}%
    \end{figure*}

   \begin{figure*}[ht!]
   \centering
   \includegraphics[width=\textwidth]{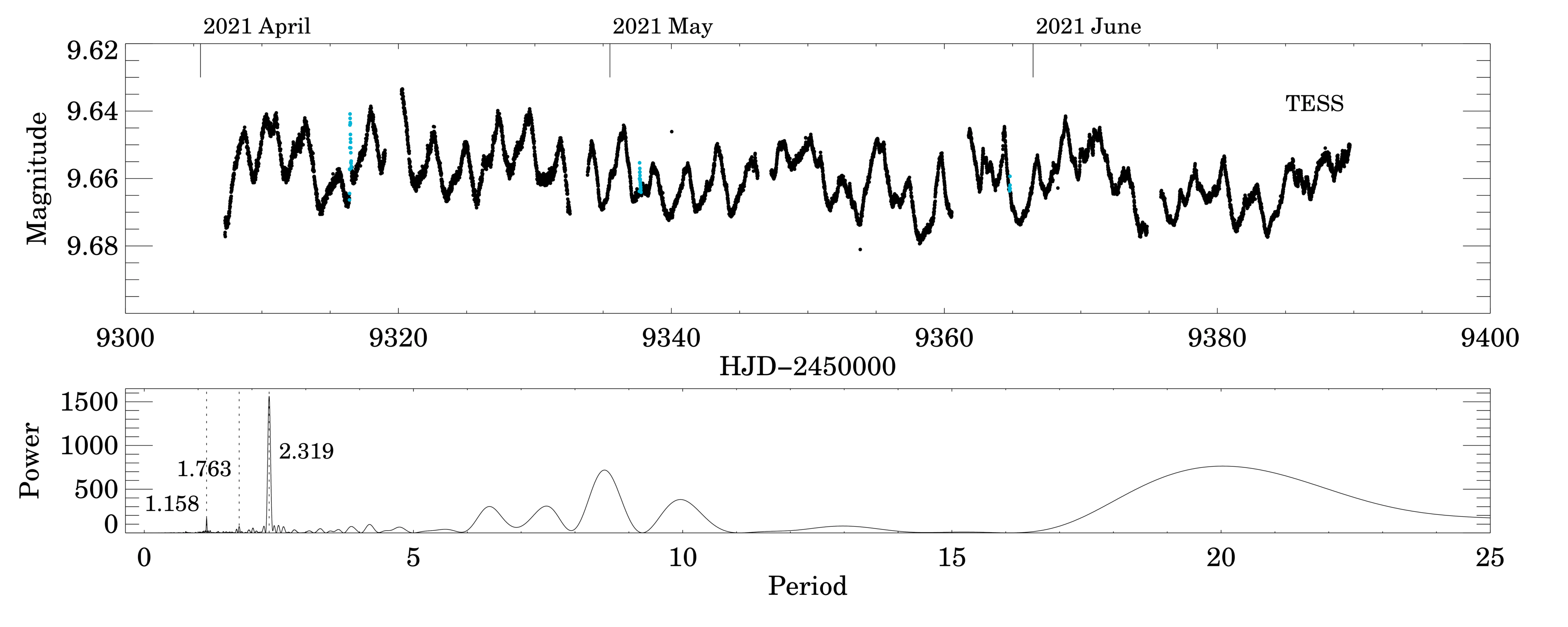}
   \caption{The TESS light curve, obtained in 2021, is indicated with a black curve, and the flare-like events are coloured in blue. The long tickmarks on the top of the figure indicate the beginning of each month. The bottom panel shows the Lomb-Scargle periodogram obtained from the 2021 TESS data.}
              \label{fig:tess_lc2021}%
    \end{figure*}

   \begin{figure*}[t]
   \centering
   \includegraphics[width=0.95\textwidth]{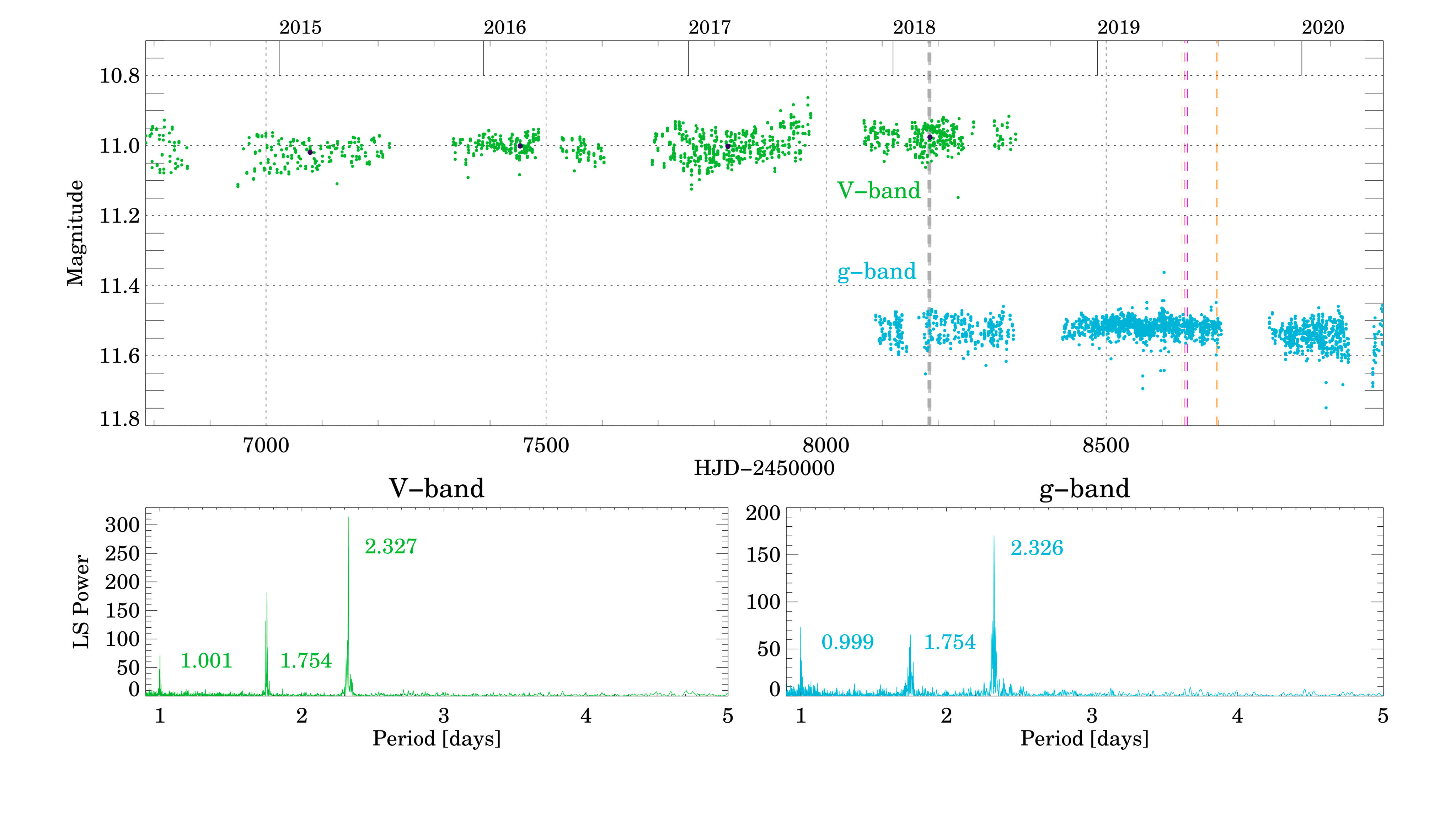}
   \caption{The ASAS-SN $V$- and $g$-band light curves are shown on the top panel along with epochs when the HARPS (gray dashed lines), the ESPRESSO (orange dashed lines) and the FEROS (purple dashed lines) were taken. The black dots indicate the median brightness for each group of data. The bottom left panel shows the Lomb-Scargle periodogram for the $V$-band data, and the left panel shows the Lomb-Scargle periodogram for the $g$-band data.}
              \label{fig:asassn_lc}%
    \end{figure*}

\subsection{2019 observing season}\label{sect::obs_2019}
The Chamaeleon I star forming region, including CR Cha, was covered in Sectors 11 and 12 of the Transiting Exoplanet Survey Satellite (TESS, \citealp{ricker2015}) in 2019. The observations started on 2019 April 22 and finished on June 19, providing almost uninterrupted broad-band optical photometric observations with 30-minutes cadence. Details on the TESS data reduction and photometry can be found in \citet{plachy2021} and \citet{pal2020}, here we only summarize the main steps. The photometry of the source was performed via differential image analysis using the \texttt{ficonv} and \texttt{fiphot} tools of the FITSH package \citep{pal2012}. This requires a reference frame, which we constructed as a median of 11 individual $64\times64$ subframes obtained close to the middle of the observing sequence.
The convolution-based approach used here exploits all of the information in the images by minimizing the difference and simultaneously correct for the various temporal aberrations (e.g. differential velocity aberration, variations in the PSF, pointing jitter corrections, etc.). A more detailed description of these methods can be found in \cite{plachy2021} or \cite{pal2020}. This method is therefore equivalent to an ensemble analysis and 
requires basically no further post-processing of the light curves such as co-trending or similar types of de-correlartion methods.
The photometry process also requires a reference flux to correct for various instrumental and intrinsic differences between the target and the reference frames. For this, we used the median of our SMARTS $I$-band photometry (see below) taken over the same time period as the TESS data. We obtained aperture photometry in the TESS images using an aperture radius of 2 pixels, and sky annulus between 5 and 10 pixels (the pixel scale is around $20''$). We note that we also inspected the most recent TESS observations, obtained between 2021 April 2 and June 24, and those data were reduced with the same procedure.

We carried out contemporaneous photometric observations with the optical-infrared imager ANDICAM mounted on the 1.3\,m telescope at Cerro Tololo (Chile) operated by the SMARTS Consortium. We obtained observations on almost every night between 2019 April 30  and June 13. We started taking images in the Johnson-Kron-Cousins $VR_CI_C$ optical and CIT/CTIO $JHK$ infrared filters, but after the first three nights, the $V$ and $R_C$ filters were unavailable for our observations, therefore, all the remaining optical images were taken with the $I_C$ filter. On each observing night we typically obtained 9 images in the $I_C$ band with an exposure time of 14\,s for a single image, and 5--7 images in the near-infrared bands with exposure times between 4 s and 15 s. For the optical images, standard bias and flatfield correction was applied by the SMARTS team. For the infrared images, dithering was performed to enable bad pixel removal and sky subtraction, which was done using custom IDL scripts. Photometric calibration in the optical was done using six comparison stars in the $6'\times6'$ field of view whose magnitudes were taken from the APASS9 catalog \citep{henden2015} and transformed to the Johnson-Cousins system using the equations of \citet{jordi2006}. For the photometric calibration in the infrared, we used 2MASS magnitudes \citep{cutri2003} of two comparison stars visible in the $2\farcm4\times2\farcm4$ field of view.

We obtained four high-resolution (R=140\,000) optical spectra with the VLT/ESPRESSO instrument as part of the DDT proposal Pr.Id.2103.C-5025, PI \'A. K\'osp\'al between 2019 May 31 and 2019 August 3, partly simultaneously with the TESS observing period. The spectra were reduced using Version 3.13.2 of the EsoReflex/ESPRESSO pipeline \citep{freudling2013}. The pipeline carries out the bias, dark, and flat field corrections, the wavelength calibration and extracts 2D spectra and merged, rebinned 1D spectra, and removes the sky lines from the spectra.
We obtained two additional spectra with the FEROS instrument (R=48\,000) on the MPEG/ESO 2.2 m telescope on 2019 June 6 and 2019 June 11. The observations were carried out in the object-calib mode, where simultaneous spectra of a ThAr lamp were taken using a comparison fiber. The spectra were reduced using a modified version of the \textsc{ferospipe} pipeline \citep{brahm2017} written in \textsc{python}. The original pipeline extracts all 33 \'echelle orders but it calibrates only 25, as it was developed to precisely measure the radial velocity. Our modified pipeline allows calibration of all 33 \'echelle orders by fitting pixel value-wavelength pairs using identified emission lines in the corresponding ThAr spectra. More detailed description of the modified pipeline can be found in \cite{nagy2021}.



   \begin{figure*}[t]
   \centering
   \includegraphics[width=0.95\textwidth]{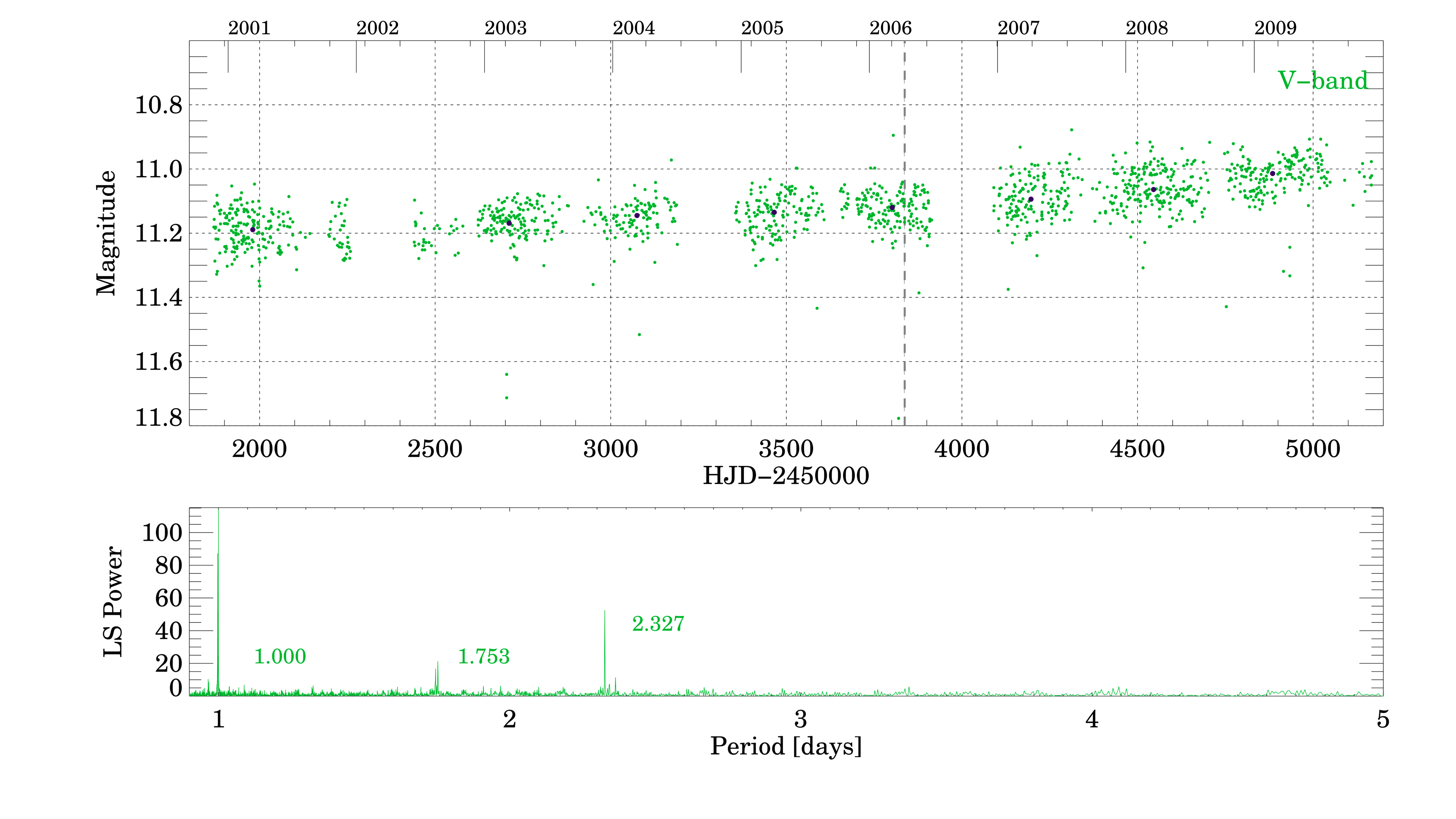}
   \caption{The top panel shows the ASAS-3 light curve with the phases when the AAT spectra were taken (vertical dashed lines), and the black dots show the median brightness for each group of data. The bottom panel shows the Lomb-Scarge periodogram for the whole ASAS-3 dataset.}
              \label{fig:asas3_lc}%
    \end{figure*}


\section{Data analysis}\label{sect::analysis}
In the following, we present the analysis of the photometric and spectroscopic data obtained in the three observing seasons.

\subsection{Analysis of the light curves}

Among the datasets available for CR Cha, the TESS light curves present the best cadence covering 56 days. 
The 2019 TESS light curve (Fig. \ref{fig:tess_lc}, middle panel), taken with 30-minute cadence, reveals both periodic and stochastic variations with a peak-to-peak amplitude of $\sim$0.04 mag. In order to find periodic signals in these photometric data, we computed the Lomb-Scargle periodogram \citep{lomb1976, scargle1982}, which is shown in the bottom panel of Fig.~\ref{fig:tess_lc}. The periodogram resulted in significant peaks at 2.314$\pm$0.033 days and 1.156$\pm$0.009 days. An analysis of the more recent TESS light curve taken with 10-minute cadence between 2021 April and June (Fig.~\ref{fig:tess_lc2021}), obtained with the same analysis technique as for the 2019 data (see Sect.~\ref{sect::obs_2019}), confirms the presence of these two peaks. The peak at $2.314$ days is consistent with the stellar rotational period found in previous studies ($P=2.3$ days, \citealt{bouvier1986}), suggesting the presence of starspots on the stellar surface. 
The $1.156$-day period is half of the stellar rotational period.
The TESS light curve also hints at long-term oscillation with a timescale of $\sim$25 days, however, the time coverage of our dataset is not long enough to probe these longer timescales.
\cite{robinson2021} modeled the effect of inclination on how periodic a light curve appears. They found that larger inclinations lead to more burst dominated light curves. The moderate 31$^{\circ}$ inclination of CR Cha indicates that the light curve is expected to be not purely periodic. Indeed, we are able to detect a rotational period in our analysis of the light curve but, at the same time, additional effects, such as accretion variation and flares also significantly contribute to the observed light curve.
The TESS light curve shows a few flare-like events as well, indicated by the blue points in the middle panel of Fig. \ref{fig:tess_lc}. During the 56-day-long TESS observing period in 2019, we found five flares with durations ranging from 3.8 hours to 7.2 hours, which gives a flare rate of 0.09 days$^{-1}$. However, these were events showing $\sim$0.02 mag brightening.

The ASAS-SN data taken between 2014 and 2020 have a sparser time coverage with respect to TESS, but are useful to learn about the long-term behavior of CR Cha. We calculated the Lomb-Scargle periodogram for all the available ASAS-SN data taken between 2014 and 2020 (Fig. \ref{fig:asassn_lc}). Since the Lomb-Scargle algorithm well suited exploration of the Fourier-spectrum also in the case of unevenly sampled data like ours, this method is able to reveal the existing periods in the ASAS-SN dataset as well. $V$-band observations are available between 2014 and 2018, and $g$-band observations are accessible between 2017 and 2020. The most significant peaks of the periodograms are at $P=2.327\pm$0.0014 in the $V$--band and $P=2.326\pm$0.0019 days in the $g$-band. These periods are in agreement with the stellar rotational period found also in the TESS light curves. Moreover, we found significant periods around 1 day in both $V$- and $g$-bands, which appear due to the fact that the observations were taken with nightly cadence. An additional 1.754-day period appears to be present in the ASAS-SN data. The TESS data from 2019 do not show signs of a period around 1.7 days, whereas a hint of this periodicity is present in the 2021 observations with TESS. We thus examined if shorter sections of the ASAS-SN data indicate its presence. We found that during the 2019 TESS observing period the ASAS-SN data also do not show any period around 1.7 days. However, randomly selected 56-day long slices of the ASAS-SN light curves show a period around 1.7 days with varying intensity. The origin of this periodic signal is unclear and surely very variable with time.

The long-term behaviour of the $V$-band light curve hints at a general, slow brightening of the system by $\sim$0.05 mag over $\sim$4~years. This effect is not observed in the $g$-band light curve, either because the effect is no longer present after 2018, or because the bluer wavelength coverage of the $g$-band filter weakens the effect.


   \begin{figure}[t]
   \centering
   \includegraphics[width=0.9\linewidth]{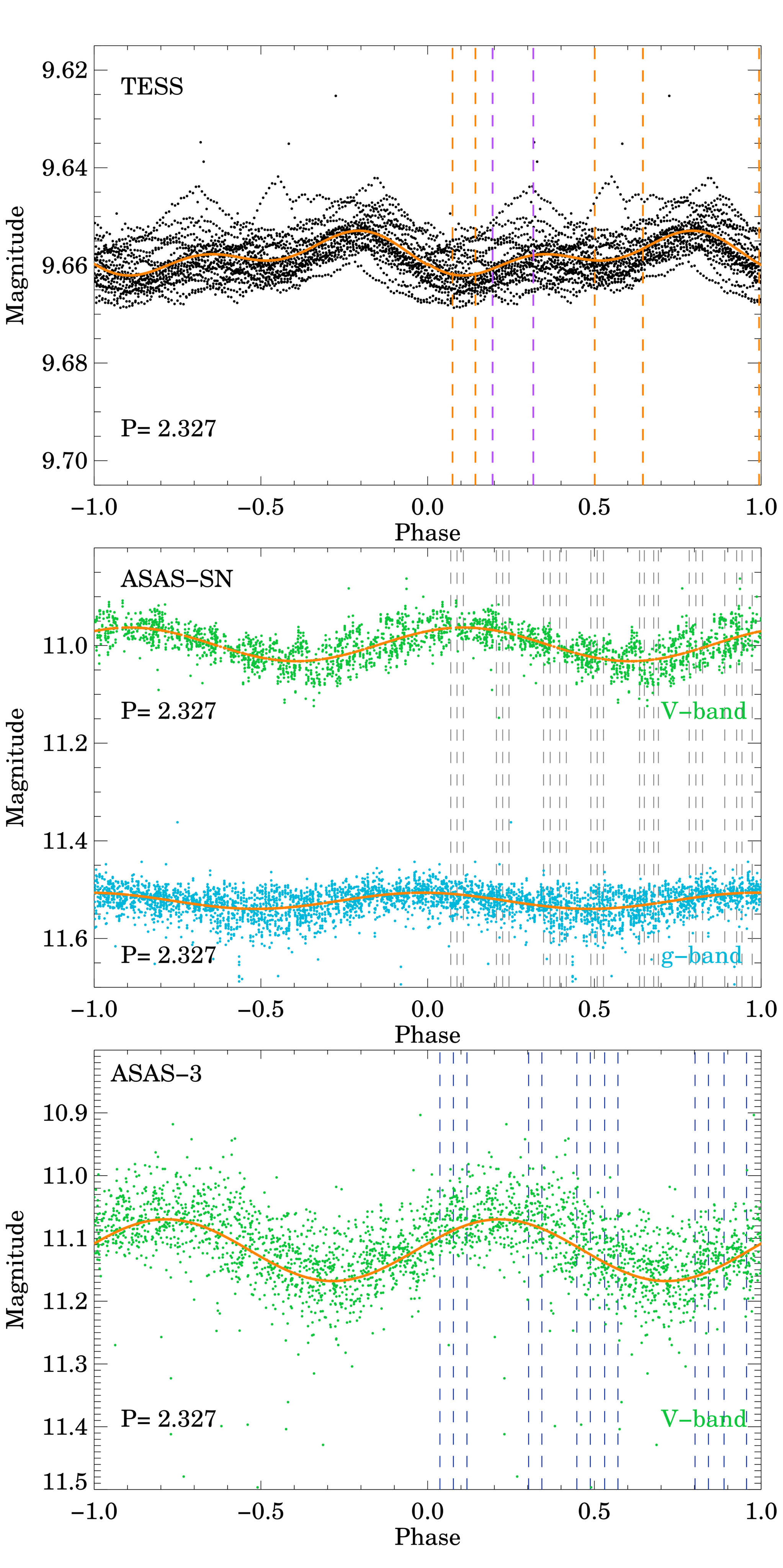}
   \caption{The phase-folded light curves of CR Cha. Vertical dashed lines indicate the epochs of the spectroscopic observations. The top panel shows the TESS measurements with the ESPRESSO (orange lines) and FEROS (purple lines) epochs, the middle panel displays the ASAS-SN observations with the HARPS epochs (gray lines), and the bottom panel presents the ASAS-3 data with the AAT epochs (dark blue lines). The orange curve shows the fitted sine curves to each data set. }
              \label{fig:crcha_phased}%
    \end{figure}
    

Similarly to the ASAS-SN data, the available ASAS-3 $V$-band photometric data, obtained between 2001 and 2009, are useful to study the long-term trends in CR~Cha. These data are shown on the top panel of Fig.~\ref{fig:asas3_lc}. We computed the Lomb-Scargle periodogram for the ASAS-3 observations, which we show in the bottom panel of Fig.~\ref{fig:asas3_lc}. We observe a well defined period at $P=2.327$ days, which is consistent with the stellar rotation period, and we interpret this signal as rotational modulation due to spots. This period is in agreement with the previous results, and with the other datasets shown here. The $2.3$ days period is present in the system for decades, and the most recent high-resolution data also confirms its presence.  On the long-term, the data reveal an increasing trend in the brightness by $\sim$0.2 mag over 8 years, thus in line with the brightening observed in the ASAS-SN data. However, no sudden, extreme brightening or dimming was detected, which indicates the stable behaviour of the system.

 As the TESS light curve reveals more stochastic variations due to its more frequent timescales sampling, and because the ASAS-SN and ASAS-3 data have longer time coverages, we adopt for our work the $P=2.327$ days period as stellar rotation period found in the ASAS-SN and ASAS-3 data for our target. We used this period to construct the phase-folded light curves (Fig.~\ref{fig:crcha_phased}). Before creating the phase-folded light curve for the ASAS-3 data, we subtracted the linear brightening trend and shifted it back to the median brightness level. 
 In addition, we also removed the long-term oscillating trend from the TESS light curve, and shifted it back to the median brightness level. We fitted a sine curve to all the different datasets, which are indicated with orange curves in Fig.~\ref{fig:crcha_phased}. The TESS light curve (top panel of Fig.~\ref{fig:crcha_phased}) reveals a trend that can be fitted better with the sum of two sine curves, indicating that multiple spots contribute to the modulation. The peak-to-peak amplitude of the fitted curves show the rotational modulation with different amplitudes in the different bands ranging from $\sim$0.009~mag in the TESS observations to $\sim$0.09~mag in the ASAS-3 measurements.


   \begin{figure}[t!]
   \centering
   \includegraphics[width=\linewidth]{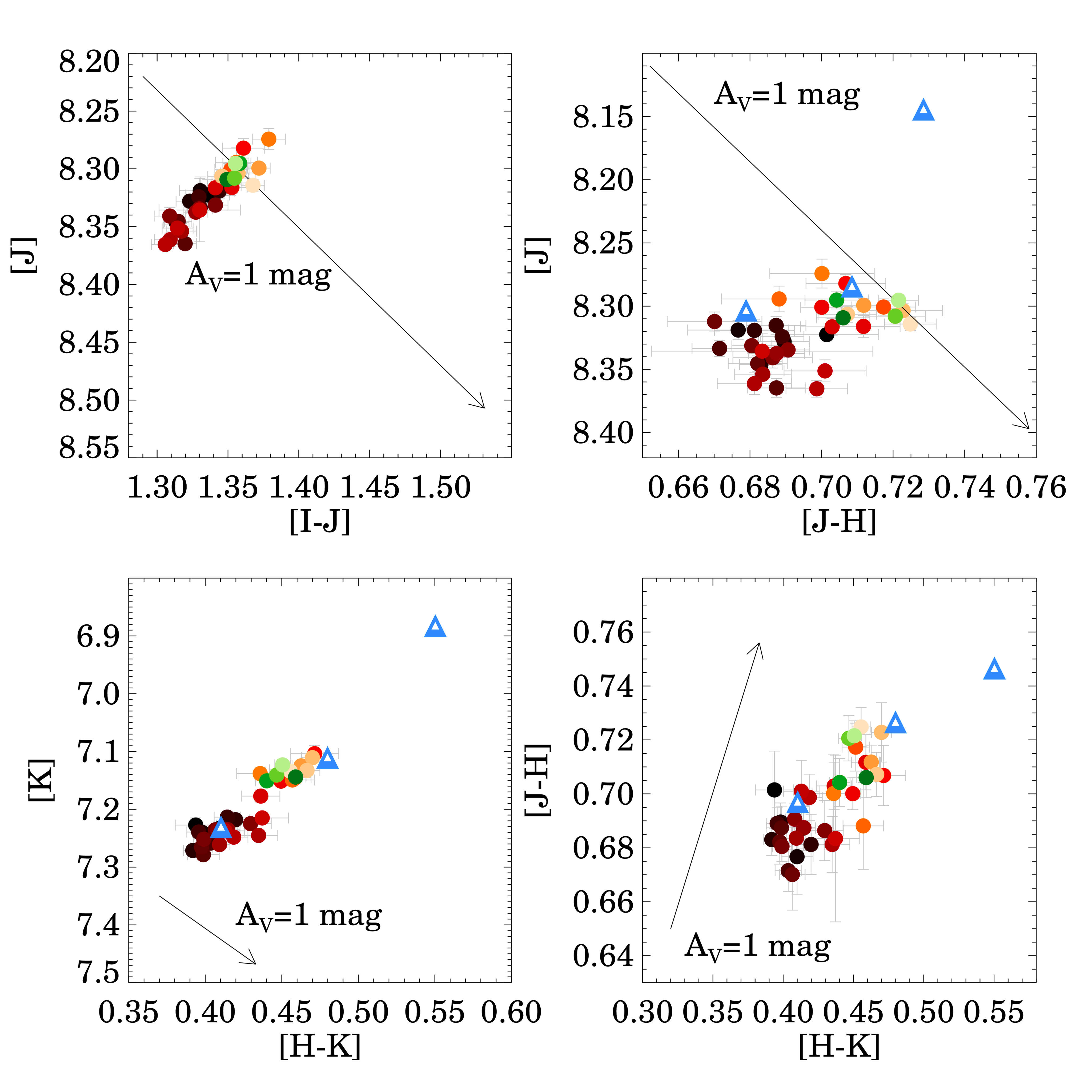}
   \caption{Color-magnitude and color-color diagrams based on the SMARTS photometry taken in 2019. The more recent measurements are indicated with lighter colors. The arrow shows a change in extinction of $A_V = 1$ mag. The blue triangles show the (rescaled) models by \citet{carpenter2001} showing how the color change for different size of the disk inner radius from 1 to 2 and to 4 $R_\odot$. }
   \label{fig:crcha_cmd}%
   \end{figure}

\subsection{Color variations}\label{sect::obs_col}

The $I_C,J,H,K$-band SMARTS observations were obtained simultaneously with the TESS observation (Fig.~\ref{fig:tess_lc}) in the 2019 observing period. For most nights, the ground-based $I_C$-band magnitudes reproduces well the space-based broad-band optical TESS photometry. The near-infrared $J,H,K$-band light curves also follow the variability seen in the TESS observations in the first 20 days, although with smaller amplitude. On the contrary, in the second half of the TESS observing window the near-infrared light curves show a brightening trend of 0.1--0.2 mag, whereas the TESS magnitudes remain constant to within 0.03 mag. Since the time coverage of the data is limited, we did not attempt to perform a periodogram analysis on the ground-based data.

In Fig.~\ref{fig:crcha_cmd}, we show the color-magnitude and the color-color diagrams obtained from our $I_C,J,H,K$-band measurements. The color-magnitude diagrams indicates that the source becomes redder as it gets brighter (negative slope). We indicated the more recent data with lighter colors in the color-magnitude diagrams, and they suggest that the system was brightening at near-infrared wavelengths during the last days of the 2019 observing period. 

The trajectories on the color-magnitude diagrams can help distinguishing different possible physical mechanisms causing the variability. These physical mechanisms include rotating starspots on the stellar surface, variable accretion, or extinction caused by the circumstellar matter.
We indicated the effect of extinction change corresponding to $A_V = 1$ mag in Fig. \ref{fig:crcha_cmd} with an arrow, which highlights that the data distribute orthogonal to the reddening vector. In addition, the presence of starspots would also result in a positive slope (redder as it gets fainter) in the color-magnitude diagrams \citep{carpenter2001}. Therefore, starspots and extinction cannot explain all of the variability characteristics we observed, which suggests the presence of an additional mechanism that affects the emission at longer wavelengths more than at shorter wavelengths. These color variations will be discussed in Sect.~\ref{sect::disc_col}.
We explored the [g - TESS] vs. TESS color-magnitude diagram, analysis of which resulted in a nearly colorless variation. We also examined if shorter-term color variations associated with rotation period appear in the color-magnitude diagrams, but we found no significant trends.

   \begin{figure*}[ht!]
   \centering
   \includegraphics[width=\textwidth]{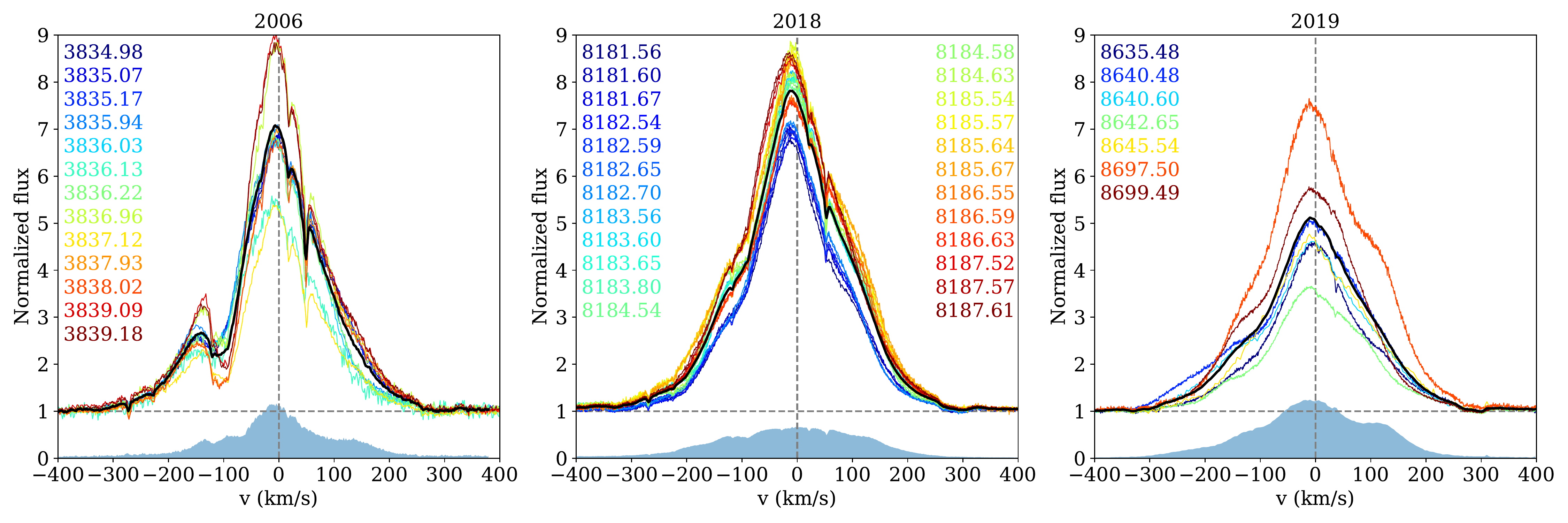}
   \caption{H$\alpha$ line profiles for all three observing seasons. The mean profiles are indicated by a thick black curve and the variance profile is shown with blue shaded area. The numbers indicate the JD of observations with the corresponding color.}
              \label{fig:variance_prof_ha}%
    \end{figure*}


   \begin{table}[h!]
   \caption[]{Equivalent widths of the H$\alpha$ lines for the 2019 campaign.}
   \label{table:ew_ha}
   $$ 
   \begin{array}{ccc}
   \hline
   \noalign{\smallskip}
   \mathrm{JD}-2450000      & EW_{H\alpha} {[\text{\AA}]} & \mathrm{Instrument}\\
   \noalign{\smallskip}
   \hline
   \noalign{\smallskip}
   8635.4841  &   -14.13  & \textrm{ESPRESSO} \\
   8640.4800  &   -17.89  & \textrm{ESPRESSO} \\
   8640.6016  &   -15.56  & \textrm{FEROS} \\
   8642.6492  &   -11.22  & \textrm{ESPRESSO} \\
   8645.5397  &   -16.38  & \textrm{FEROS} \\
   8697.5039  &   -29.56  & \textrm{ESPRESSO} \\
   8699.4947  &   -19.30  & \textrm{ESPRESSO} \\
   \noalign{\smallskip}
   \hline
   \end{array}
   $$ 
   \end{table}

\subsection{Spectroscopy}

   \begin{figure*}[t!]
   \centering
   \includegraphics[width=0.36\textwidth]{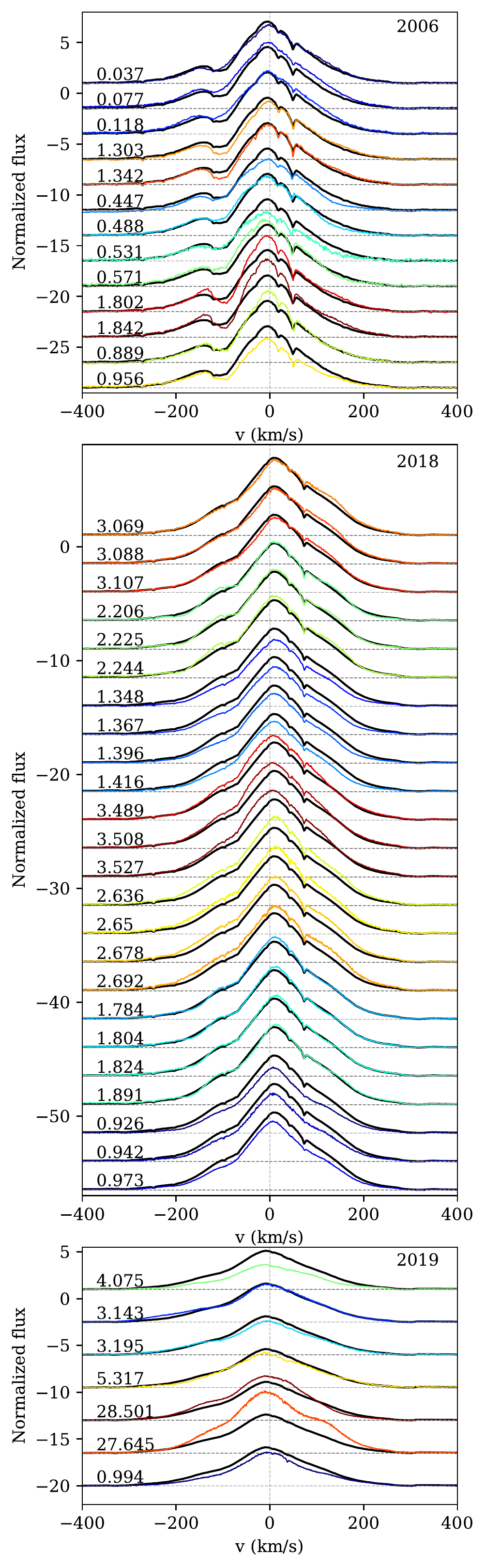}
   \hspace{0.12\textwidth}
   \includegraphics[width=0.36\linewidth]{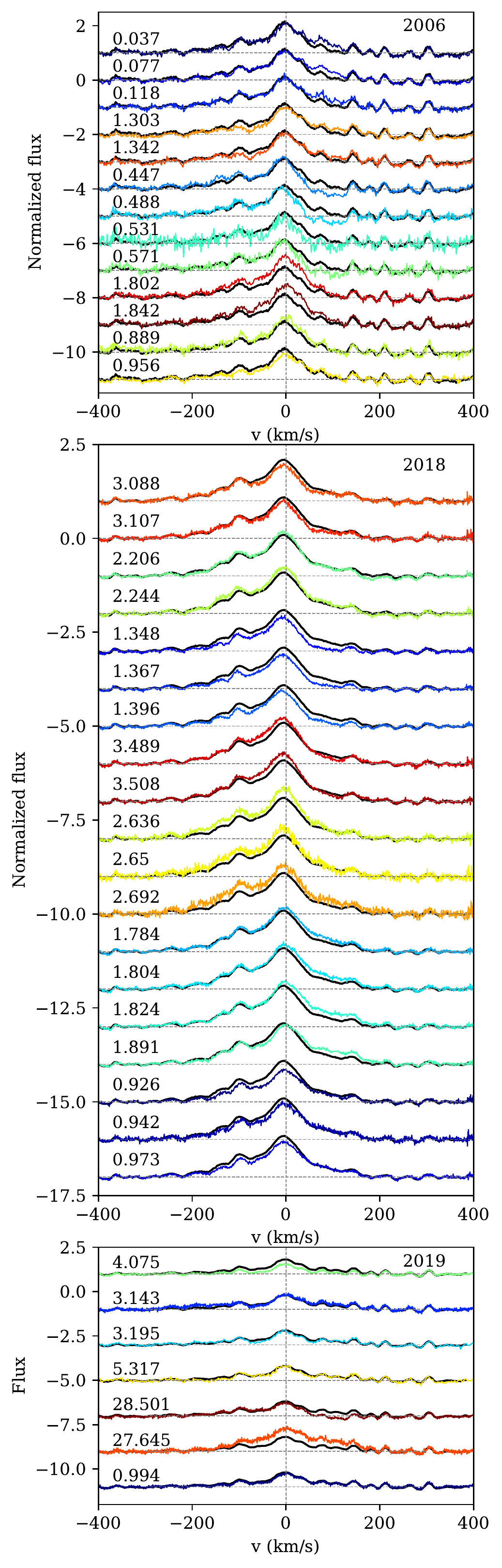}

   \caption{The left column shows the H$\alpha$ spectra ordered by rotation phase, and the right column displays the H$\beta$ lines ordered by rotation phase for each observing season. The top panels show the AAT, the middle panels show the HARPS, and the bottom panels show the ESPRESSO and FEROS measurements. The phases with the rotational number are indicated on the left side of each line. The thick black lines indicate the mean line profile for each observing season.}
              \label{fig:crcha_ha_phase}%
   \end{figure*}

   \begin{figure*}[t!]
   \centering
   \hspace{-4mm}
   \includegraphics[width=0.34\linewidth]{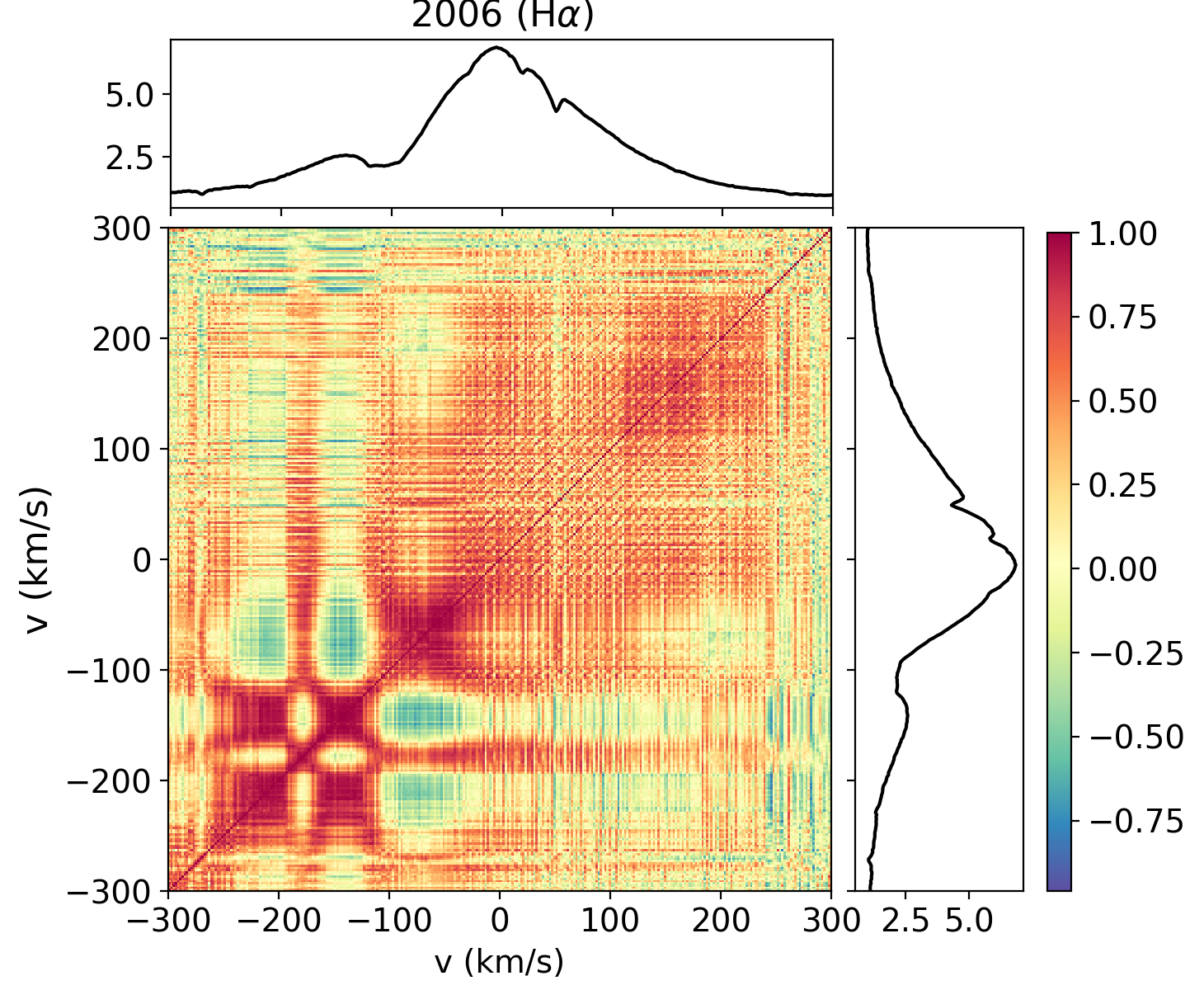}
   \hspace{-2.5mm}
   \includegraphics[width=0.34\textwidth]{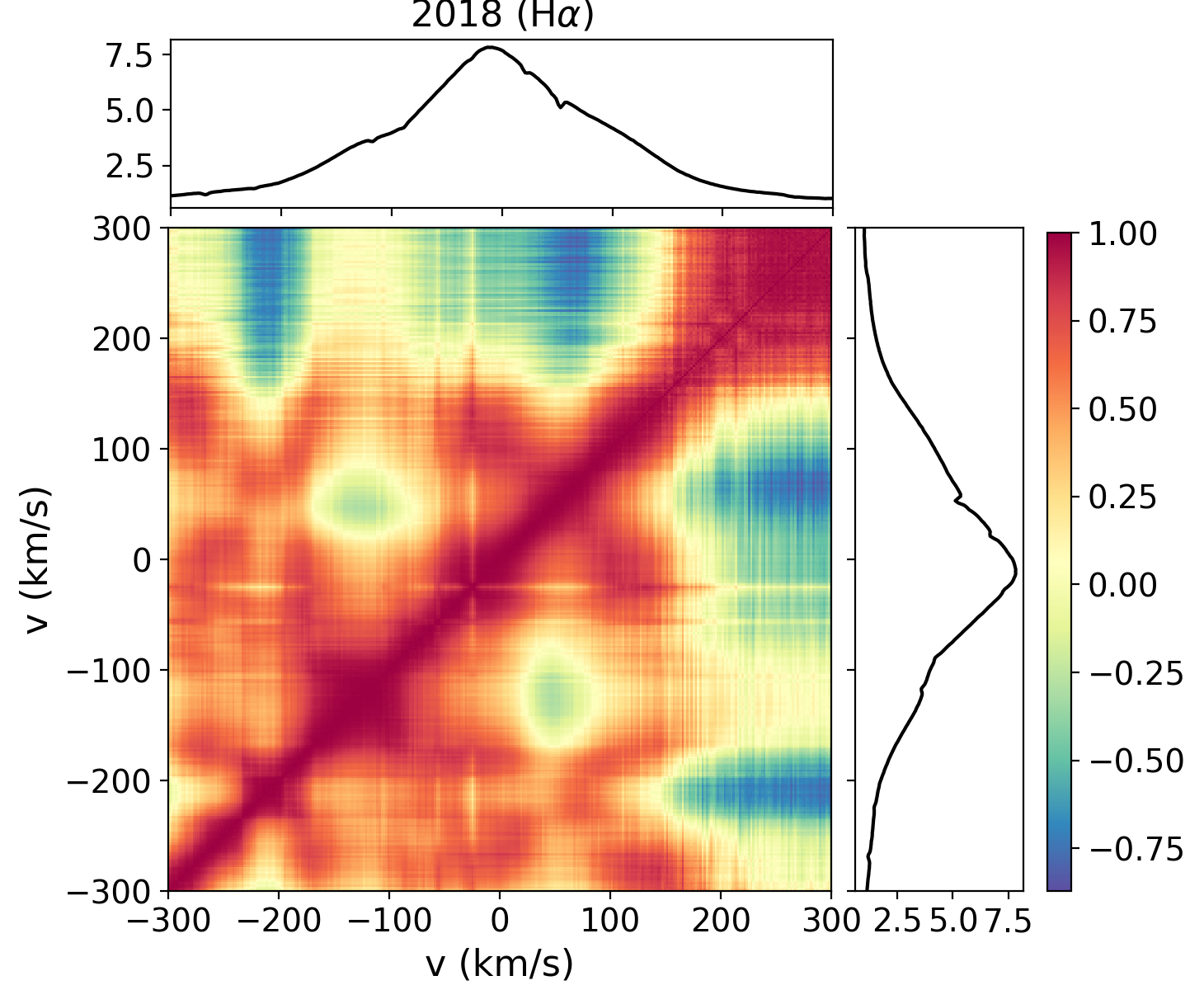}
   \hspace{-3mm}
   \includegraphics[width=0.32\textwidth]{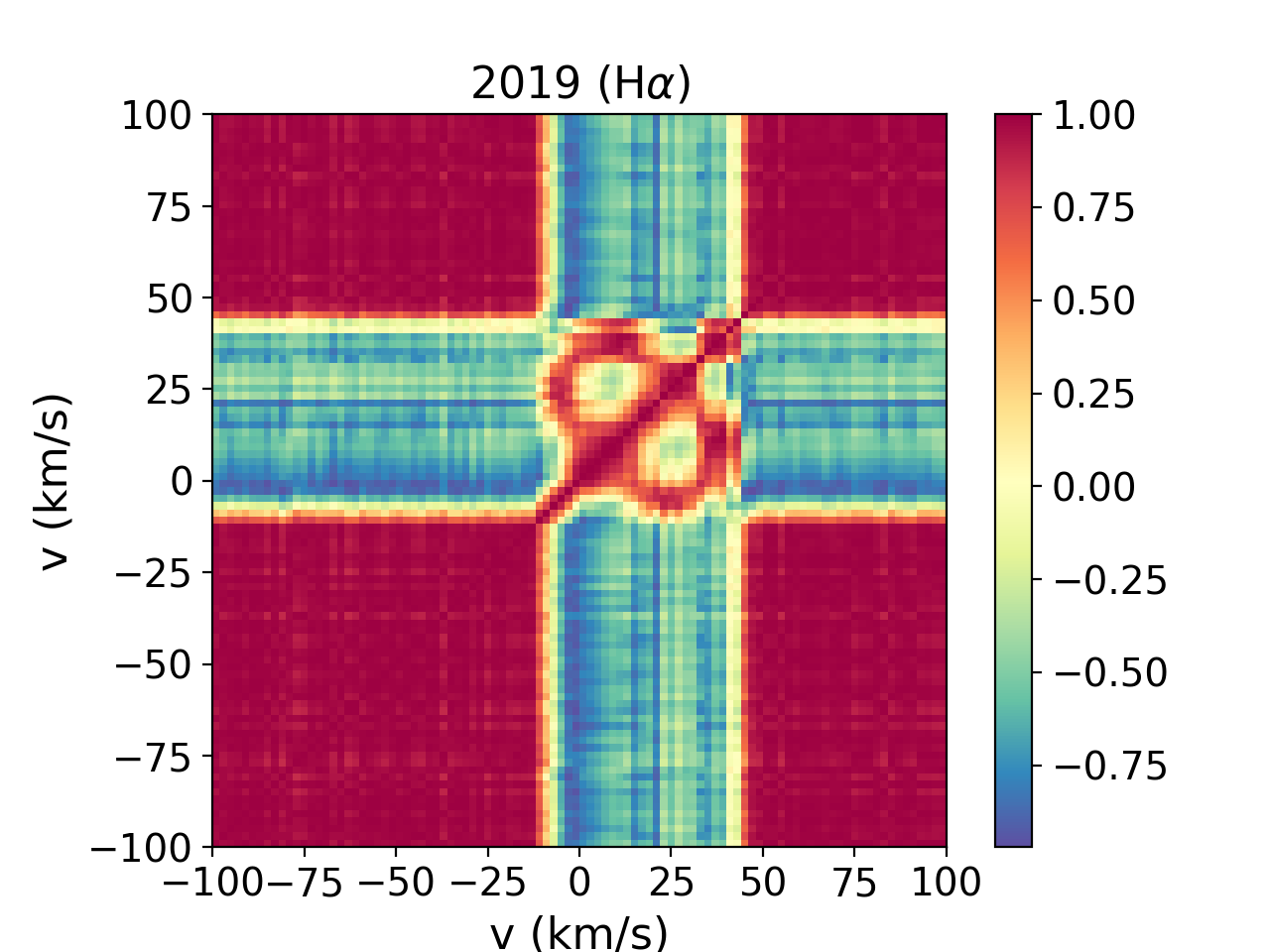}
   \caption{Correlation matrices for the H$\alpha$ lines. The positive values (more red color) indicates correlation, the negative values (bluer color) show anti correlation, and the values around zero (yellow color) indicate no correlation. Left: 2006 (AAT), Middle: 2018 (HARPS), Right: 2019 (ESPRESSO and FEROS)}
              \label{fig:corr_mtx_ha}%
   \end{figure*}

       \begin{figure}[ht!]
   \centering
   \includegraphics[width=0.8\linewidth]{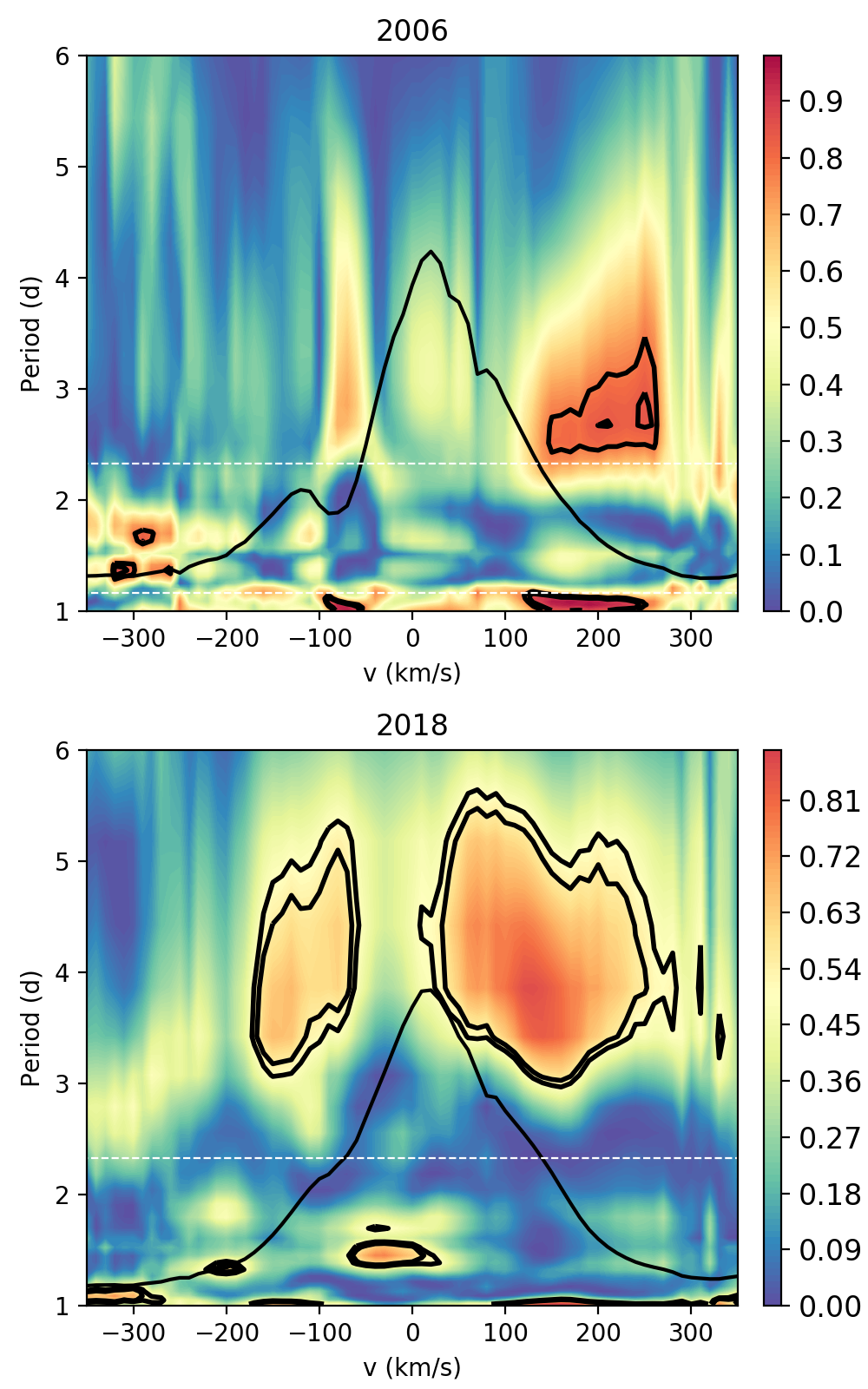}
   \caption{Two dimensional periodogram for the H$\alpha$ lines. The colorbar represents the periodogram power varying from zero (blue) to the maximum power (red). The inner contours correspond to the 95\% confidence level, the outer contours correspond to the 85\% confidence level. The stellar rotational period is marked by the horizontal white dashed line.}
              \label{fig:2dperiodogram_ha}%
    \end{figure}

In order to display the phase coverage of the spectroscopic observations, we indicated in Fig.~\ref{fig:crcha_phased} with vertical dashed lines the phases at which the spectra used in this work were taken. The middle panel of the figure shows that the HARPS data from 2018 cover the whole phase-space. On the other hand, the AAT observations from 2006 do not cover the dimmest phase (bottom panel), and the ESPRESSO and FEROS measurements from 2019 are more sparse in the phase coverage (top panel). We note that, despite the fact that two ESPRESSO observations were taken outside of the TESS observing period, we plotted them over the top panel in order to display the phase coverage of all observations.    

The shape of the emission line profiles present in the spectra of CR Cha change throughout the observing seasons on yearly, monthly and daily timescales. 
The strongest emission line in all spectra is the H$\alpha$ line. In addition, we detected the H$\beta$ and the [O~I] 630 nm lines in weaker emission. The [O~I] 557 nm emission line present in the HARPS and AAT spectra is not astronomical but due to emission in our atmosphere. Indeed, this line was removed from the ESPRESSO and the FEROS spectra during the data reduction process. 
Besides the emission lines, we detected the Na~I doublet and the Li~I 670.8 nm absorption lines in all spectra. The Ca~II infrared triplet region (CaIRT) is covered only by the FEROS spectra. Since one line of the triplet is located between two orders, only two CaIRT lines were detected, and they appear in weak absorption. In addition, the Ca~II H and K lines appear in emission in the ESPRESSO and FEROS spectra but they are not covered by the AAT observations. The Ca~II K line was detected in the HARPS spectra as well, however, the noise level is too high to identify the Ca~II H line.

In the following, we analyze the variability of the two permitted emission lines detected in the spectra, and of the veiling of the photospheric lines.

\subsubsection{Analysis of the H$\alpha$ line}

The H$\alpha$ line profile shows both short term and long term variability. The mean H$\alpha$ line profiles for each observing season, indicated with thick black lines in Fig.~\ref{fig:variance_prof_ha}, show that the general H$\alpha$ line profile has changed remarkably on yearly timescales. 

All H$\alpha$ line profiles show a strong central component, however, the intensity of this component varies over time. The largest change in the amplitude, including the weakest peaks, appears during the 2019 observing season. However, it should be noted that this observing season is more sparsely sampled but covers a longer period than the earlier ones: the sampling probes the monthly timescales. Besides the amplitude variations, the H$\alpha$ line exhibits morphological variations. A red bump appears at about 100 km/s on the night with the strongest emission, moreover, the entire blue wing becomes relatively strong during one of our observations.
In contrast, the 2006 data display different line profiles with respect to the more recent datasets. The central peak reaches a maximum among our spectra, but also varies by a factor of 1.5 in peak intensity. A strong blueshifted absorption feature is present around $-100$ km/s, and this line profile shape is preserved during the entire 2006 observing period with varying amplitudes.
The 2018 data show very stable line profiles during the entire HARPS observing period with a small amplitude variation. A small blue absorption appears here as well, however, the profiles are more symmetric than in 2019.

In order to examine the temporal variance of the line profiles in each velocity bin, we calculated the variance profile as described in \cite{johns1995}. We show the variance profiles in Fig.~\ref{fig:variance_prof_ha} as the blue shaded areas. The variance profiles indicate the strongest variability of the lines at the central peak for all three observing periods, with the largest amplitude variations in 2019. This is likely caused by the varying accretion onto the star. The 2018 data show only small amplitude changes with little variations of the amplitude with velocity. In contrast, the 2006 observations indicate variability at the dip around $-100$ km/s as well as changes in the amplitude of the central peak, which might indicate a varying wind component. Finally, the 2019 data show strong variations both in the central peak, as well as at $+100$ km/s and $-100$ km/s. While the latter could be again ascribed to a varying wind, the former is possibly the result of varying accretion geometry.

We show the H$\alpha$ line profiles ordered by phase for all observations in Fig.~\ref{fig:crcha_ha_phase} along with the median profile indicated with black. The 2006 observations on the top panel show that the blueshifted absorption around $-100$ km/s becomes more pronounced from phase 0 to 0.3, then the strength of component approaches the level of median profile towards the photometric minimum. As the light curve comes out of a minimum, the absorption gets more pronounced again. The redshifted excess emission compared to the median profile around 150 km/s weakens as the phase progresses, and it is weakest right before the minimum. After the photometric minimum, this component strengthens again.

We show the H$\alpha$ line profiles ordered by phase for the 2018 observations in the middle panel of Fig.~\ref{fig:crcha_ha_phase}. In this case, the blueshifted bump at $-150$ km/s becomes stronger from phase 0 to phase 0.2. Towards the photometric minimum, this component gets weaker than the median value, however, it becomes stronger again closest to the minimum. As the light curve leaves the minimum, this component weakens again. At the same time, the central peak weakens from phase 0 to phase 0.4, however, it also becomes stronger closest to the minimum, and weakens again as the light curve leaves the minimum.

The bottom panel of Fig.~\ref{fig:crcha_ha_phase} shows the 2019 ESPRESSO and FEROS observations ordered by phase. Since these observations cover a longer time span, they are less sensitive to the variability on short timescales, however, they can reveal the variability on monthly timescales. This shows that the amplitude of central component of the H$\alpha$ line is highly variable during our observing period, which suggests changing accretion onto the central star.

The observed emission lines consist of multiple components and display variability on various timescales. These components might originate from different physical processes. In order to examine the relationship between the variations of these components, we calculated correlation matrices \citep{johns1995}. We divided each emission line into small velocity intervals and calculated the correlation coefficient ($r$) between every pair of velocity bins as follows:

$$r_{ij} = \frac{\sum (x_i - \bar{x})(y_i - \bar{y})}{\sqrt{\sum (x_i - \bar{x})^2 \sum(y_i - \bar{y})^2}},$$
where $\textbf{x}$ and $\textbf{y}$ contains the intensity values of each epoch for a chosen velocity bin. Calculating the correlation coefficient between every pair of velocity bin results in a correlation matrix.

The auto-correlation matrices for the H$\alpha$ lines are shown in Fig.~\ref{fig:corr_mtx_ha}. The 2006 observations show correlation in the blue wing of the H$\alpha$ line, between around -150 km/s and -250 km/s. There is also  anticorrelation between around -80 km/s and -250 km/s, and -80 km/s and -150 km/s. The pattern is slightly different for the 2018 observations: anticorrelation appears between the red and the blue parts of the line, at $\sim$50 km/s and -120 km/s. The ESPRESSO and FEROS observations resulted in very different auto-correlation matrix. This probably results from the larger temporal gap between the individual observation.

In order to examine whether the photometry and the spectroscopy are linked in a way that lines vary with the same period as the photometric brightness, we applied a Lomb-Scargle periodogram analysis in each velocity bin for the H$\alpha$ and the H$\beta$ lines. The resulting 2D periodograms for the 2006 and the 2018 observing seasons are plotted in Figure~\ref{fig:2dperiodogram_ha}. We drew contours around the regions corresponding to the 85\% and 95\% confidence levels.

During the 2006 observing season, the red wing of the H$\alpha$ line and the blueshifted absorption component ($\sim$100 km/s) showed significant periodic behaviour with a period of $\sim$1.1 days, about half of the stellar rotation period. As the data cover 4.24 days, which is less than the twice the rotational period, we might not be sensitive to variability on timescales of the rotation period. However, a $\sim$2--3-days period was found in the redshifted wing of the H$\alpha$ line with 85\% confidence level. In contrast, the 2018 HARPS observations reveal a somewhat longer periodicity ($\sim$3--5 days) in the redsihfted side between 0 and 250 km/s and on the blueshifted side between -100 and -200 km/s, and do not seem to be modulated by the rotational period. 

\subsubsection{Analysis of the H$\beta$ line}

\begin{figure*}[ht!]
   \centering
   \includegraphics[width=\textwidth]{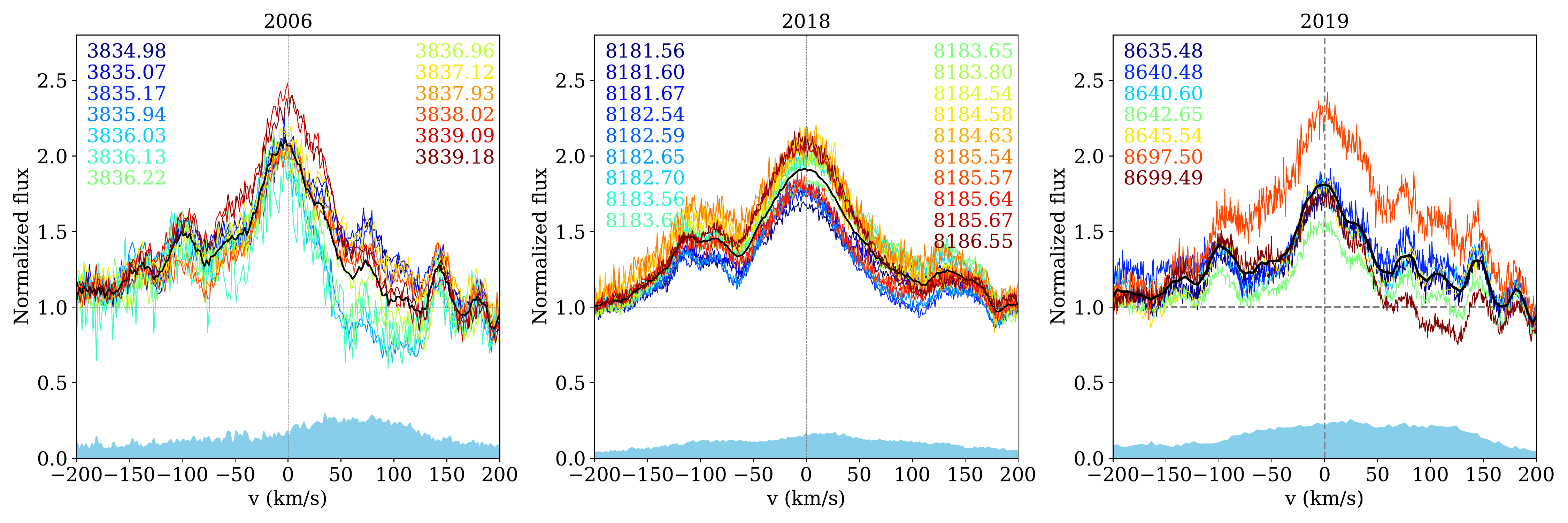}
   \caption{H$\beta$ line profiles for all three observing seasons. The mean profiles are indicated by a  thick black curve and the variance profile is shown with blue shaded area.}
              \label{fig:variance_prof_hb}%
    \end{figure*}

The profile of the H$\beta$ line also displays variability throughout the years, however, during a given observing period, it shows variability with small amplitude. We calculated the normalized variance profiles as described above, and indicated them with blue shaded area in Fig.~\ref{fig:variance_prof_hb}.

There are several photospheric absorption lines in the H$\beta$ region which are superposed on the H$\beta$ emission. In order to eliminate the effects of the absorption lines, we subtracted the rotationally broadened spectrum of $\epsilon$ Eri, which has the same spectral type as CR~Cha, from the observed spectra, and this process resulted in more defined peaks. The results show (Fig.~\ref{fig:variance_prof_hb}) that most of the variability come from the amplitude changes of the line during all observing seasons. On the other hand, the strength of the mean line profile slightly changes throughout the years. Overall, they do not show as strong variability as the H$\alpha$ line. We present the H$\beta$ line profiles ordered by phase in Fig.~\ref{fig:crcha_ha_phase}. The results show similar pattern in the amplitude variations as the H$\alpha$ line, however, no significant morphological variations were found.

\begin{figure}[ht!]
   \centering
   \includegraphics[width=0.8\linewidth]{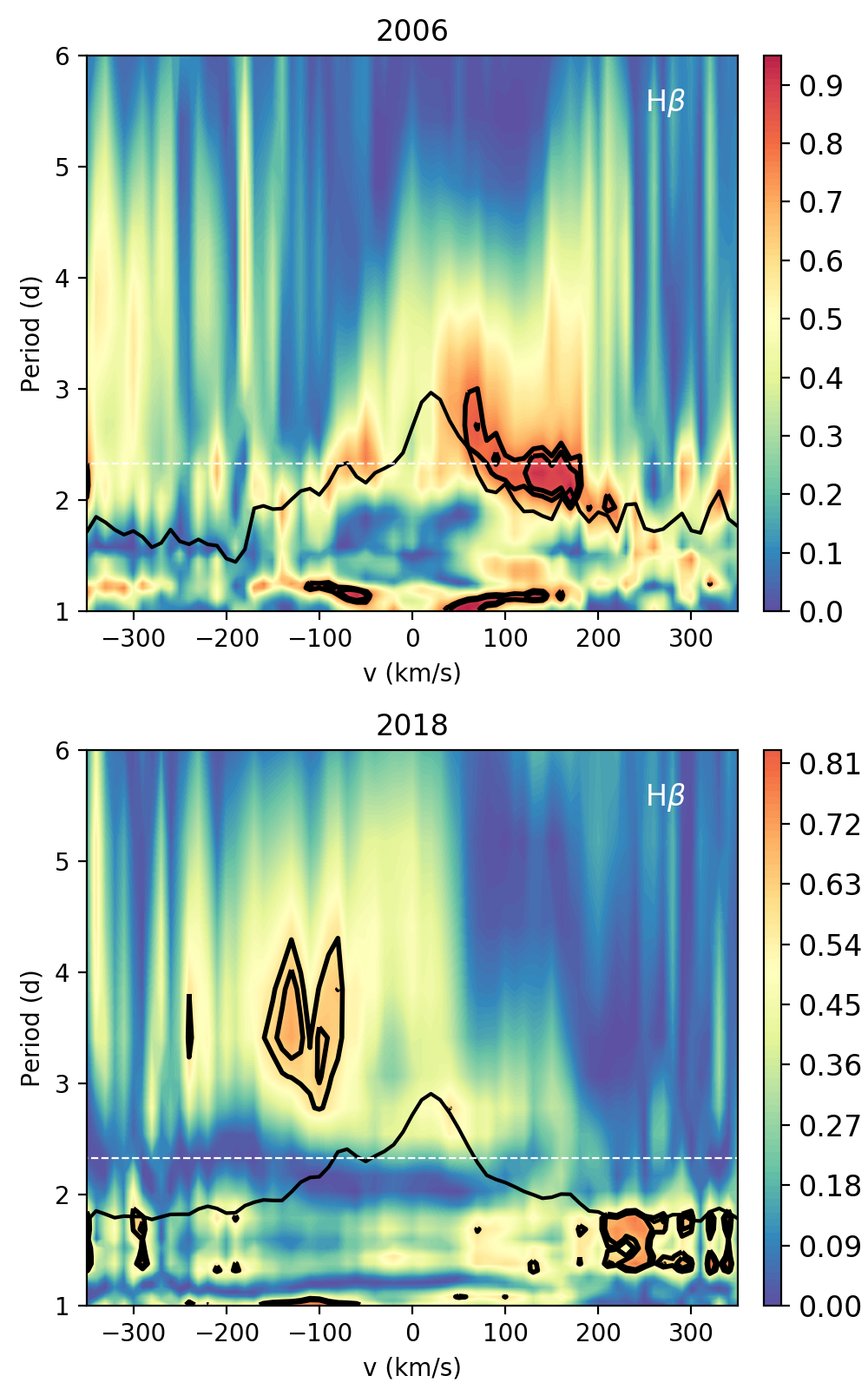}
   \caption{Two dimensional periodogram for the H$\beta$ lines. The colorbar represents the periodogram power and the inner contours correspond to the confidence levels (as in Figure \ref{fig:2dperiodogram_ha}).}
              \label{fig:2dperiodogram_hb}%
    \end{figure}

The 2D periodograms for the H$\beta$ line are shown in Fig.~\ref{fig:2dperiodogram_hb}. The 2006 measurements reveal a period around the rotation period in the red wing of the line, but the 2018 data shows only a period around  $\sim$3--4 days, similarly to H$\alpha$ but with smaller extent in terms of velocity.




  \begin{figure}[ht!]
   \centering
   \includegraphics[width=\linewidth]{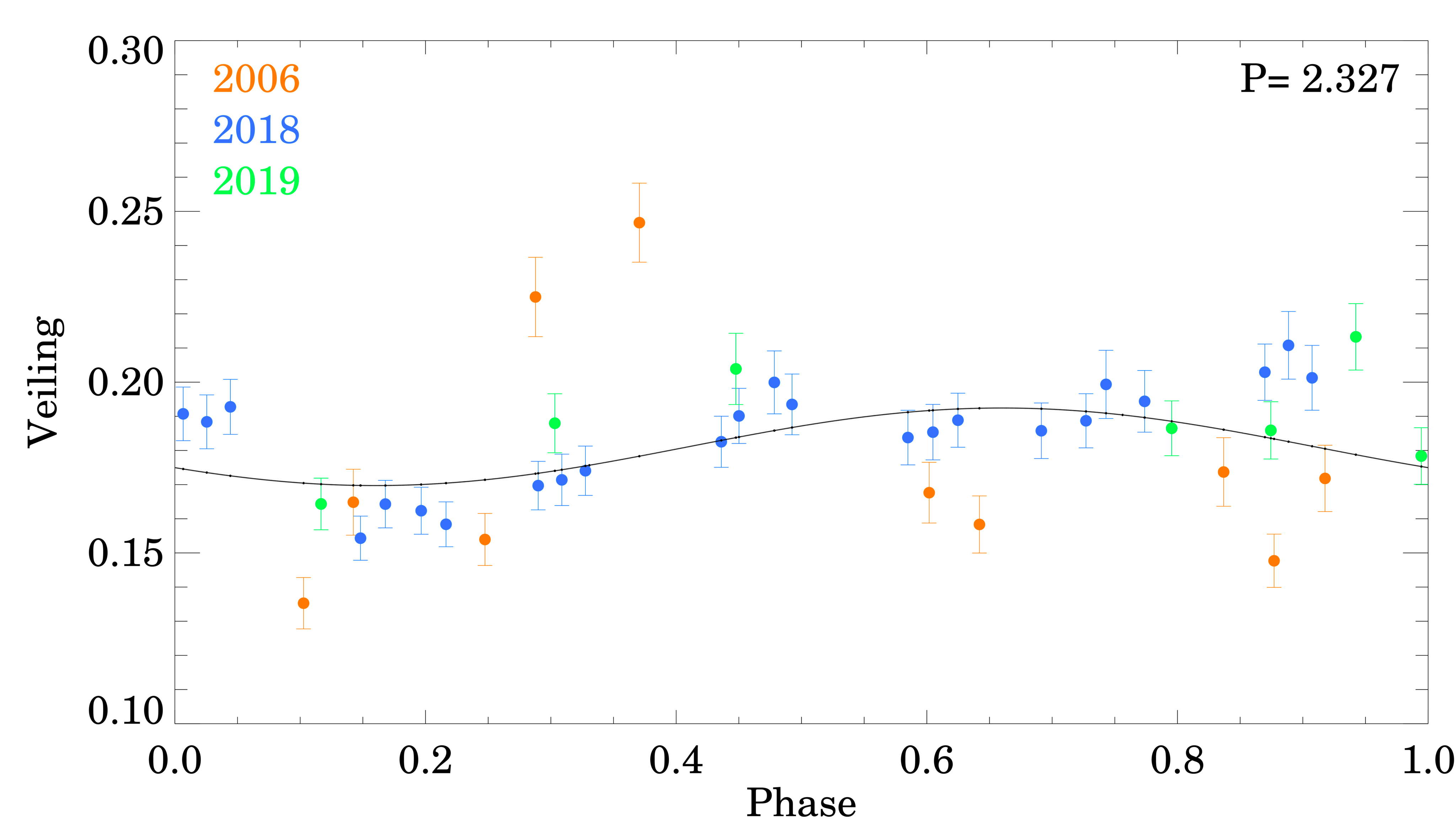}
   \caption{Veiling variations during the three observing seasons as a function of orbital phase, using the P=2.327 days period obtained from our period analysis of the light curves.}
              \label{fig:veilig_phase}%
    \end{figure}

\subsubsection{Veiling measurements}

The photospheric spectrum of classical T Tauri stars is often veiled by additional continuum emission. The veiling continuum is attributed to the shock-heated gas that results in a hot spot on the stellar surface below the accretion stream. The veiling of a star can be determined by measuring the equivalent widths of photospheric lines

$$W^0_{eq} = W_{eq} (V+1), $$
where $W_{eq}$ is the measured equivalent width, $W^0_{eq}$ is the reference equivalent width, and $V$ is the veiling.

Instead of using only a single absorption line, we constructed an average absorption line profile using the Least Squares Deconvolution (LSD) method \citep{donati1997}. We used the unpolarized LSD profiles for calculating the veiling, however, since veiling is wavelength dependent, we recalculated the LSD profiles for a smaller region of the spectrum between 530 nm and 700 nm. As $\epsilon$ Eri is a non-accreting star with the same spectral type as CR Cha, we used the rotationally broadened spectrum of $\epsilon$ Eri as a reference spectrum in order to calculate the absolute veiling of CR Cha. CR Cha shows moderate veiling of $\sim$ 0.2 with small variations. As veiling is often considered as a measure of the accretion rate, the small veiling variations hint at a modest change in the accretion rate by a factor of $\sim$3.

We also examined whether the veiling, which was obtained from the absorption lines, varies in phase with the rotation period. Figure \ref{fig:veilig_phase} shows that the majority of the veiling values are almost constant and they are only slightly modulated by the rotational period. However, we note that our results show three additional outlying data points in terms of the veiling at phases 0.33 and 0.76, which are out of the range of Fig.~\ref{fig:veilig_phase}.


\section{Discussion}\label{sect::discussion}

CR Cha shows small amplitude yet significant photometric variability on timescales from hours to years with a peak-to-peak amplitude ranging from $\sim$0.05 mag in the optical TESS observations to $\sim$0.15 mag in the $K$-band data. Periodic behavior was found in both new TESS observations from 2019 and 2021 and archive ASAS-SN (2014-2020) and ASAS-3 (2000-2010) catalogues. This confirmed that the $\sim$2.3-day periodicity is present in the system for decades. This periodicity can be attributed to stellar rotation caused by spots of the stellar surface. The brightness variations can be modelled with simple spot models, but the available colors and the uncertainty on each value do not allow us to distinguish between cold (T$\sim$4350 K) or hot (T$\sim$5300 K) spots. It is likely that both exist on the surface of this star, which is known to host a complex magnetic field \citep{hussain2009}. In addition, the ASAS-3 and the ASAS-SN V-band observations hint at a general, slow brightening of the system: the ASAS-3 data show $\sim$0.2 mag brightening during $\sim$8 years of observations, and the ASAS-SN V-band light curve indicates a brightening of $\sim$0.05 mag over $\sim$4 years. The origin of this brightening is unclear, however, change in the average spot coverage as spots slowly evolve over time might contribute to the observed behavior effect.
On the other hand, the TESS data, thanks to their much higher photometric accuracy, reveal a more stochastic behavior with peak-to-peak amplitude of $\sim$0.05 mag, probably due to accretion variability, and a few flares. The multifilter $J$, $H$ and $K$-band light curves have similar shapes to the TESS light curve, however, the amplitude of the variability is different. The $J$-band data have the smallest amplitude of $\sim$0.09 mag and the amplitude increases towards the $K$-band, which have $\sim$0.17 mag peak-to-peak amplitude. While the TESS data probe the variability on the shortest timescales, the ground-based photometry reveals a trend beyond the rotational modulation. 

The most striking variable features in the spectra are the highly variable Balmer emission lines. The mean H$\alpha$ line profiles show strong variations in both intensity and line profile on yearly timescale. This suggests that different physical mechanisms contribute to the observed variability. In this section we discuss how we interpret these spectroscopic and photometric variations.

\subsection{Line morphology variations}


We discuss here the variability of the emission line profiles during the three observing periods.
The observed shifts in the variability period on decadal timescales might originate from the formation of multiple hot spots on the stellar surface at different times: one dominant hot spot would cause modulation at the stellar rotation period, while multiple hot spots at different latitudes would result in different variability periods. Multiple hot spots are expected on CR Cha, as the star hosts a complex magnetic field \citep{hussain2009}. As the 2019 observations do not have as good phase coverage as the previous epochs, we cannot check if there is any periodicity.

In comparison with studies focused on other T Tauri systems, we see in this target that the period analysis of the H$\alpha$ lines often do not exhibit a single and highly significant peak. As the H$\alpha$ line is expected to form not only at the accretion spot but also in the accretion funnel and in the stellar winds, the variability of this line is also influenced by the variations of the circumstellar environment. However, the rotation period might appear in the periodograms with varying strengths. 
It is instructive to compare our findings with similar analyses of individual targets. 
In the LkCa 15 system, the strength of the emission at the line center is modulated by the stellar rotation, and this effect is clearly displayed in the periodogram \citep{alencar2018}. On the other hand, the inverse P Cygni profile of the H$\alpha$ line of the HQ Tau system displays modulation on the timescale of the stellar rotation not only in the line center but all the way to the red wing \citep{pouilly2020}. Another interesting system to mention is V2129 Oph. When analyzing this target, \cite{sousa2021} compared their results with the periodograms for the same system analyzed by \cite{alencar2012}, and they found some differences on $\sim$decadal timescale, similarly to our results. \cite{sousa2021} discuss a few possible explanations for the observed phenomena, such as variations in the magnetic field strength, which would result in variations in the magnetospheric truncation radius, or latitudinal differential rotation, or a more complex magnetic structure with two major funnel flows originating at different radii in the inner disk.
However, in the case of V2129 Oph, a periodicity longer than the stellar rotation period is present at the line center for a time of almost a decade. The likely cause of the periodicity is a structure that is present beyond the corotation radius which is stable for almost a decade. In the case of CR Cha, the rotation period has been stable for decades but we see that the line profiles (and also the photometry) do vary with longer periods as well, likely due to variations of the circumstellar environment.



The overall H$\alpha$ line profile variations show different characteristics during the three observing seasons indicating the significance of different variable physical mechanisms on yearly timescale. The 2018 observations show the least amount of variability, mostly showing modulations presumably due to changes in the accretion rate. The 2006 observations reveal slightly different line profile with an additional blueshifted absorption component, which is usually associated with ejection processes, such as wind. In addition, the amplitude variations of the central peak are larger than the ones during the HARPS observations in 2018, which suggests larger change in the accretion rate. The 2019 observations show line profiles similar to those measured in 2018, however, based on their variance profile, these data show the largest amount of variability due to the changing accretion, similarly to the 2018 observing season. 

The H$\beta$ line profiles do not show as prominent morphological changes as the H$\alpha$ lines. However, the amplitude of the H$\beta$ lines vary both over the observing seasons and within one set of observations. The 2018 data show the least amount of variability, similarly to the H$\alpha$ line, indicating very small changes in the accretion rate. The 2019 and 2006 data exhibit comparable and larger amount of variability across the whole line, which hints at slightly larger changes in the accretion rate. We quantify these variations in the next Section.


\subsection{Timescales of accretion variability}

\begin{figure*}[ht!]
   \centering
   \includegraphics[width=\textwidth]{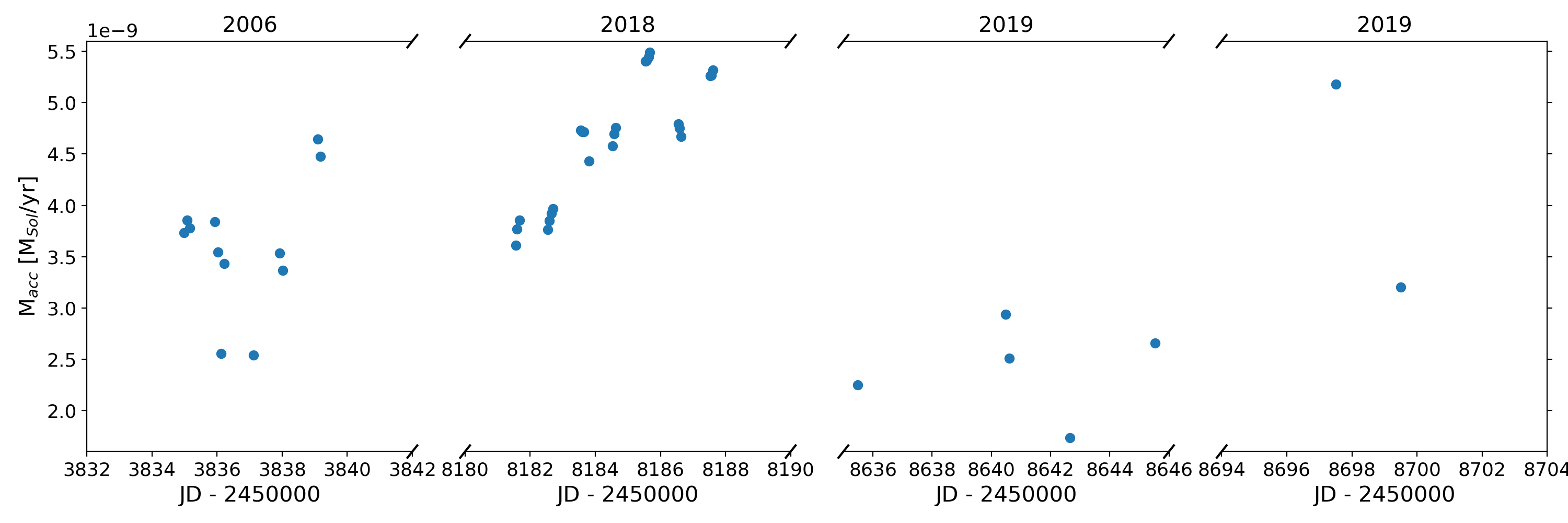}
   \caption{Mass accretion rates calculated at each observing epoch from the estimated luminosity of the H$\alpha$ line for CR Cha.}
              \label{fig:mdot_time}%
    \end{figure*}

After describing how the observed emission lines vary in morphology with time, we want to investigate here the timescales on which these variations are largest. Our data allow us to explore the timescales from hours to a decade in one particular object. We calculated the equivalent width of the H$\alpha$ line at each epoch, and since our photometric observations indicate that the brightness of the star barely changes, the equivalent width should be proportional to flux variations. In order to confirm that the variations in equivalent width roughly correspond to variations in accretion rates, we converted the equivalent width values to accretion rates for each observation with the following procedure. First, we calculated the H$\alpha$ line luminosity using the continuum flux value around the H$\alpha$ line measured on the flux-calibrated X-Shooter spectra of this target in 2010 \citep{manara2016}. 
As the mean brightness of the system did not change significantly between 2010 and 2019 (see Figures~\ref{fig:asassn_lc}-\ref{fig:asas3_lc}), we used the same continuum flux value for both the 2018 and the 2019 observations, whereas we accounted for the $\sim$0.1 mag dimmer state during the 2006 by decreasing the continuum flux by the corresponding flux ratio. The line luminosity was then converted into accretion rate using the empirical relation from \cite{alcala2017}. 
Finally, we compute \macc \, with the classical relation \citep{hartmann2016}. The values of \macc \, at the time of our observations are shown in Fig.~\ref{fig:mdot_time}, and fall between $\sim$2-5$\cdot 10^{-9} M_\odot$/yr, in line with previous results \citep[e.g.,][]{manara2016,manara2019}.

\begin{figure}[ht!]
   \centering
   \includegraphics[width=0.9\linewidth]{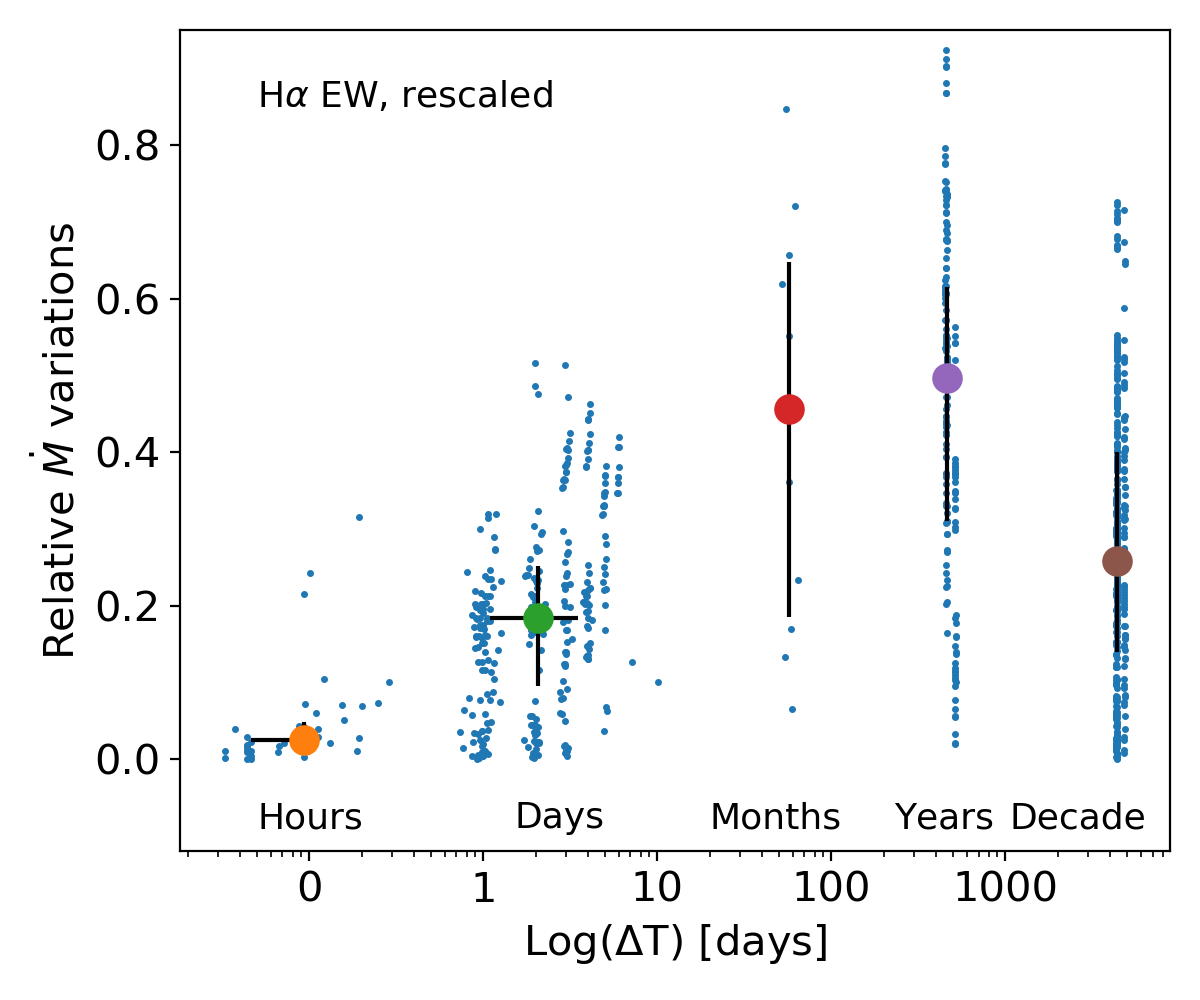}
   \caption{Relative variations in \macc, computed from the estimated luminosity of the H$\alpha$ line, on timescales from hours to 13 years.}
              \label{fig:timescales_macc}%
    \end{figure}

We then obtained the relative amplitude of the accretion rate variations compared to the mean accretion rate value. In order to cover all the possible timescales from our dataset, every value of the accretion rate was compared with the ones obtained from every other spectrum. These are displayed with small blue dots in Fig. \ref{fig:timescales_macc}. We also marked the median amplitude of the variations for the different timescales with larger colored dots, and indicated the 0.25 to the 0.75 percentile of the distribution of the points with black lines. Our results show that the amplitude of the variations increases from hours to several days, and it saturates when in the range between a week to a month, which is not well sampled by our data. This result is partially in agreement with the conclusions of \citet{costigan2014}. Indeed, the comparison of this result with the stellar rotation period suggests that the maximum of the variations are on timescales that are longer than the rotation period, whereas \citet{costigan2014} suggested that the maximum variability is on timescales of the order of the rotation period. In addition, there is an apparent drop at the timescales of a decade. However, this effect might arise from the fact that our sample on decadal timescale is less rich in data than on shorter timescales.

Studies of large samples of low-mass young stars show that the the amplitudes of variability in the accretion rate range from small fractions to $\sim$1 order of magnitude, with typical value of $\sim$0.4 dex \citep{venuti2014}. In this context, the accretion rate of CR~Cha is smaller than the typical value for young stars with mass >1$M_{\odot}$ but the accretion rate variations of CR~Cha reaches the typical value presented in \cite{venuti2014}. The typical timescales of accretion variability of classical T~Tauri stars has also been explored in previous works in the framework of large surveys. \cite{nguyen2009} examined the spectra of several low-mass pre-main-sequence stars and found that the amplitude of the variations increases from hours to several days, after which it saturates. \cite{costigan2012} analyzed the optical spectra of 25 targets in the Chamaeleon~I region and derived an upper limit on the dominant timescale of the accretion variability: they suggest that observations on timescales of a few-weeks are sufficient to characterise the majority of the accretion-related variations in typical young stars over a period of $\sim$1 year. \cite{costigan2014} studied 15 T Tauri and Herbig Ae stars over a wide range of timescales and found that the majority of the variations occur as gradual changes in the H$\alpha$ emission and the period of days is the dominant timescale of these variations. Our data confirms that accretion variability has a maximum of intensity on weeks to months timescales.

\subsection{Explanation of the color variations}\label{sect::disc_col}

As discussed in Sect.~\ref{sect::obs_col}, the data we have obtained allow us to study the variations in the near-infrared colors for CR~Cha in 2019. We compare these variations with the models developed by \citet{carpenter2001}. 

 \cite{carpenter2001} examined the near-infrared color variations of young stars and described three physical models as the possible origin of the observed color changes. In the first of their models, the light variations are caused by starspots, which modulate the brightness of a star as stellar rotation alters the fractional spot coverage. This model resulted in either nearly colorless fluctuations or positive slope on the color-magnitude diagram (i.e. the object becomes bluer as it gets brighter), depending on the fractional spot coverage and spot temperature. This is not observed for CR~Cha. 
 The extinction model attributes the observed color changes to extinction, which result from inhomogeneities in either the inner circumstellar environment or the molecular cloud that move across the line of sight. The extinction vectors can be calculated from the interstellar reddening law, and they also result in a positive slope on the color-magnitude diagram. However, the extinction slope is shallower by $25^{\circ}$ than those expected from hot spots. Again, this is not what is observed for our target. 
 
 The third model that \cite{carpenter2001} consider is an accretion disk model, which takes into account the emission from a circumstellar disk as a source of near-infrared variability. The near-infrared variability may originate either from changes in the mass accretion rate or changes in the inner disk structure that alter the amount of absorbed and reprocessed stellar radiation. Variations in the accretion rate or in the structure of the inner disk result in a negative slope in the near-infrared color-magnitude diagram.
This negative slope (target becomes redder as it gets brighter) on the color-magnitude diagrams we observe in CR~Cha (Fig.~\ref{fig:crcha_cmd}) is peculiar, and typically observed in a small fraction of targets \citep[$\sim$1\%,][]{carpenter2001}.
This behavior suggests that the color variations are not due to changing extinction or to the presence of stellar spots, but more consistent with the accretion disk model described by \citet{carpenter2001}.

Based on our available data, we can attempt to test the hypothesis that the variations in \macc \ are the cause of the color variability. In order to test this hypothesis, 
we compared the equivalent widths of the H$\alpha$ lines in the 2019 observing epoch (Table~\ref{table:ew_ha}) with the data points on the color-magnitude diagrams which were taken on the same nights (indicated with green color in Fig.~\ref{fig:crcha_cmd}). Unfortunately, as these data points cover only a small range on the diagram, it is difficult to recognise any pattern. However, a rough comparison is possible.  
The disk model in \citet{carpenter2001} indicates that the expected photometric variations can be a few tenth of magnitudes for accretion rates changing from $\dot{M}\sim3\cdot10^{-9} M_{\odot} \rm /yr^{-1}$ to $\dot{M}=10^{-7}M_{\odot} \rm yr^{-1}$. The comparison between these models and our photometric data suggests that, if the color variations are only due to accretion rate variations, the variability of the accretion rate during our observations were smaller than a factor of $\sim$5. This is in line with equivalent width variations, which indicate that the the accretion rate varies by a factor of $\sim$3. Thus, the possibility that the color variations are due to an accretion rate variation is compatible with our data.


On the other hand, variations in the size of the disk inner hole, or variations in the thickness of the inner edge of a warped disk may also result in changes in the reprocessed radiation \citep[e.g.][]{carpenter2001}. Their models suggest a few tenth of magnitude variation in the near-infrared bands when the disk inner hole size changes between $1 R_{\odot}$ and $4 R_{\odot}$.
CR~Cha shows less than one-tenth of a magnitude variation in Fig.~\ref{fig:crcha_cmd}, indicating a smaller change in the inner hole size. We indicated in Fig.~\ref{fig:crcha_cmd} the model values of the color and the magnitude changes from \citet{carpenter2001} with respect to the bluest datapoint of our observations on each panel. 
We see that our data line up with two of the open triangles, which correspond to a change in the inner hole sizes from 4 to 2\,$R_{\odot}$ in the case of mass accretion rate of $\dot{M}\sim3\cdot 10^{-9} M_{\odot} \rm yr^{-1}$, which is in line with the measured value for CR~Cha. Therefore, variations in the structure of the inner disk are also in line with the observations. 

\cite{roquette2020} carried out an investigation of the near-infrared color variations of a large sample of young stars using $J$, $H$ and $K$- band observations. They calculated the trajectories in the color-space, and measured slopes for stars describing linear trajectories. They found that among the 144 measured $\frac{\Delta K}{\Delta (H-K)}$ slopes 115 systems had negative slopes. On the other hand, among the 196 measured $\frac{\Delta J}{\Delta (J-H)}$ slopes only 3 systems had negative slopes. This might imply that amplitudes of variability due to variable accretion rate and changes in the inner disk are more significant for the $J$ magnitude and $J-H$ color than for the $K$ magnitude and $H-K$ color. However, the flux in the $J$ band is dominated mainly by the stellar flux, whereas the amount of light re-emitted by the disk in the near-IR is larger for longer wavelengths. All in all, \cite{roquette2020} still found 1.7 times more stars with variability caused by extinction/spot than caused by changes in the inner disk or accretion, which places CR Cha in the class of the rarer systems.


We cannot unfortunately measure the size of the inner disk hole with our data. However, the circularly polarised spectra from our 2006 AAT and 2018 HARPS observations allow us to derive the expected magnetic field truncation radius ($R_{\rm tr}$). \cite{hussain2009} showed that CR Cha hosts a complex magnetic field including dipole and octuple magnetic field components as well. As the dipole component is expected to truncate the disk, while the octupole dominates on a smaller scale and at the surface of the star \citep{gregory2008}, we determined the magnetospheric truncation radius using the dipole component. Assuming a dipole field configuration, the magnetic field truncation radius can be calculated using the following relation \citep{bouvier2007}:
$$\frac{R_{\rm tr}}{R_{\ast}} = \frac{B_{\ast}^{4/7} R_{\ast}^{5/7}}{\dot{M}^{2/7}(2GM_{\ast})^{1/7}}$$
We caution that this relation to derive the truncation radius may not be applicable to stars hosting complex multipolar fields. With this caveat, we can calculate $R_{\rm tr}$ for CR Cha. According to \citet{hussain2009}, the dipole component of the large-scale magnetic field is $\sim$100 G; this results in $R_{\rm tr} = 2.9\ R_{\star}$, which is equivalent to 7.3 $R_\odot$, assuming  $R_{\star}=2.5\ R_\odot$ \citep{hussain2009}.
Fig.~\ref{fig:crcha_cmd}, depicting the 2019 data, indicates that the change in the inner disk hole size is from 4\,$R_\odot$ to 1\,$R_\odot$. Ignoring variations in \macc, on which $R_{\rm tr}$ has a shallow dependence, such variation would imply a variation in $B_\star$ of 24\,G. Unfortunately, our data cannot test this variations. Future studies should aim to cover several rotational periods with spectropolarimetric data and near-infrared photometry to test whether variations in the magnetic field strength, and thus in the truncation radius, are compatible with the observed color variations. For now, we can only conclude that the variations in the near-infrared colors are compatible with the observed variations in \macc, but we cannot exclude that variations in the inner disk structure also contribute to these color variations.


\section{Conclusions}\label{sect::concluions}
We presented a spectroscopic and photometric study of the classical T Tauri star CR Cha.  We examined several datasets from three different observing seasons in 2006, 2018 and 2019. Based on the analysis of the photometric data, we found that a 2.326 days period is present on the dataset, which is in agreement with the stellar rotation perion. This indicates that the photometric variations are modulated by the stellar rotation period due to the presence of stellar spots. In addition, the TESS light curves revealed a few flare like events thanks to its high temporal resolution.
By comparing the color changes of the $I$, $J$, $H$ and $K$ band observations with the models of \citet{carpenter2001}, we found that the color variations are due to changes in the inner disk properties. These can be due to variations in the mass accretion rates compatible with the ones we observe, or with changes in the inner hole size of the circumstellar disk.

The presence of the hydrogen Balmer lines allowed us to examine the accretion process in more detail. The H$\alpha$ line shows the largest variability in shape in 2019, whereas it showed only small amplitude variations in 2018. The comparison between the three observing seasons indicates significant morphological changes, but the variations of the central peak strengths due to changes in the accretion were the most significant differences in all observing seasons.

Our extensive dataset allowed us to examine the variability on timescales from minutes to decades. The photometric data can be interpreted as variations on timescales of $\sim$2.3 days due to stellar rotation. On a decadal timescale, we found a slight brightening trend, and on the shortest timescales of minutes-hours, flare-like ($\sim$0.02 mag) events also cause short brightenings. 

The H$\alpha$ line amplitude, thus accretion rate changes, show fluctuations on wide range of timescales. The amplitude of the variability increases from minutes to several days, and it saturates when it reaches the weeks-month timescale. Our results suggest that a week or so is the dominant timescales of accretion variability for this target. This is in line with previous works, and suggest that, apart from secular variability, any measurement of accretion is likely to vary by $\sim$0.4 dex on a timescale of weeks to months.

Future works covering various timescales of variability with spectroscopy should aim at covering the phase of the periodicity of the target at all observing seasons, which would provide a possibility of checking whether variations in accretion are due to variations in the magnetic field topology. CR Cha belongs to the rare class of targets that gets bluer when fainter  at infrared wavelengths. This phenomenon should be investigated with surveys, in order to verify whether the size of the inner disk varies with time, as suggested by the models. In the brightest of these systems it would be important to assess if this behaviour is predominantly associated with a particular type of stellar magnetic field (e.g. complex, variable multipolar fields or more stable simple axisymmetric fields).

\begin{acknowledgements}
G.Zs. has received support from the ESO Studentship Program Europe. 
This project has received funding from the European Research Council (ERC) under the European Union's Horizon 2020 research and innovation programme under grant agreement No 716155 (SACCRED).
This project has received funding from the European Union's Horizon 2020 research and innovation programme under the Marie Sklodowska-Curie grant agreement No 823823 (DUSTBUSTERS).
This research received financial support from the project PRIN-INAF 2019
This work was partly supported by the Deutsche Forschungs-Gemeinschaft (DFG, German Research Foundation) - Ref no. FOR 2634/1 TE 1024/1-1.
"Spectroscopically Tracing the Disk Dispersal Evolution".
The research leading to these results has received funding from the LP2018-7 Lend\"ulet grants of the Hungarian Academy of Sciences.
G.Zs. is supported by the \'UNKP20-3 New National Excellence Program of the Ministry for Innovation and Technology from the source of the National Research, Development and Innovation Fund.
This work has received  financial support of the Hungarian National Research, Development and Innovation Office – NKFIH Grant K-138962.
We thank K. Vida for the service in the spot modelling.

\end{acknowledgements}


%

%

%
%


\end{document}